\documentclass[12pt,a4paper]{article}
\usepackage{jheppub}
\usepackage{bbm}
\usepackage{graphicx}
\usepackage[utf8]{inputenc}
\usepackage[T1]{fontenc}

\usepackage{tikz}

\usepackage{amsmath,amssymb,amsfonts,bm,amscd,amsthm}
\usepackage{array}
\usepackage{enumitem} 
\usepackage{tikz-cd}                     
\def \Z{\mathbb{Z}}
\def \R{\mathbb{R}}
\def \C{\mathbb{C}}
\def \CP{\mathbb{CP}}
\def \Spin{\mathrm{Spin}}
\def \CN{\mathcal{N}}

\theoremstyle{plain}
\newtheorem{exercise}{Exercise}
\font\manual=manfnt

\newcommand{\bi}{\begin{itemize}}
\newcommand{\ei}{\end{itemize}}

\newcommand{\bea}{\begin{eqnarray}}
\newcommand{\eea}{\end{eqnarray}}
\newcommand{\be}{\begin{equation}}
\newcommand{\ee}{\end{equation}}
\newcommand{\ben}{\begin{eqnarray*}}
\newcommand{\een}{\end{eqnarray*}}
\newcommand{\bem}{\begin{pmatrix}}
\newcommand{\eem}{\end{pmatrix}}
\newcommand{\bl}{\begin{align}}
\newcommand{\el}{\end{align}}
\newcommand{\beg}{\begin{gather}}
\newcommand{\eeg}{\end{gather}}

\newenvironment{myenumerate}{
\begin{enumerate}
   \setlength{\itemsep}{1pt}
   \setlength{\parskip}{0pt}
   \setlength{\parsep}{0pt}}{\end{enumerate}}
\newenvironment{myitemize}{
\begin{itemize}
   \setlength{\itemsep}{1pt}
   \setlength{\parskip}{0pt}
   \setlength{\parsep}{0pt}}{\end{itemize}}

\def\={\;  = \;}

\def\wh{\widehat}

\newcommand{\TrH[1]}{ {\raise -.5em
                      \hbox{$\buildrel {\textstyle  {\rm Tr } }\over
{\scriptscriptstyle \cH _ {#1}}$}~}}
\newcommand{\TrS}{ {\raise -.5em
                      \hbox{$\buildrel {\textstyle  {\rm Tr } }\over
{\scriptscriptstyle SP}$}~}}
\newcommand{\TrSb}{ {\raise -.5em
                      \hbox{$\buildrel {\textstyle  {\rm Tr } }\over
{\scriptscriptstyle SP_{b}}$}~}}
\newcommand{\TrSs}{ {\raise -.5em
                      \hbox{$\buildrel {\textstyle  {\rm Tr } }\over
{\scriptscriptstyle SP_{s}}$}~}}

\newcommand{\res[1]}{{\raise -.5em 
\hbox{$\buildrel{\textstyle{\rm Res}}\over {\scriptscriptstyle {#1}}$}}}
\newcommand{\tends[1]}{{\raise -.5em 
\hbox{$\buildrel{\longrightarrow}\over {\scriptscriptstyle {#1}}$}}}

\newcommand{\cA}{\mathcal{A}}

\newcommand{\cC}{\mathcal{C}}

\newcommand{\cH}{\mathcal{H}}

\newcommand{\cN}{\mathcal{N}}
\newcommand{\CO}{\mathcal{O}}

\newcommand{\bQ}{\ensuremath{\mathbb{Q}}}

\newcommand{\bZ}{\ensuremath{\mathbb{Z}}}

\newcommand{\scM}{\ensuremath{\mathcal{M}}}
\newcommand{\scN}{\ensuremath{\mathcal{N}}}

\newcommand{\hh}{{\widehat h}}

\newcommand{\IZ}{\mathbb{Z}}
\newcommand{\IH}{\mathbb{H}}

\newcommand{\Zint}{\mathbb{Z}}

\renewcommand{\b}{\beta}

\newcommand{\g}{\gamma}

\def\iimg{ { i}}

\renewcommand{\l}{\lambda}
\newcommand{\m}{\mu}

\newcommand{\s}{\sigma}                                   
\renewcommand{\t}{\tau}

\newcommand{\D}{\Delta}

\newcommand{\G}{\Gamma}

\newcommand{\vf}{\varphi}

\newcommand{\vth}{\vartheta}

\newcommand{\1}{{\textbf{1}}}
\newcommand{\inn}{{\,\in\,}}

\newcommand{\half}{\frac{1}{2}}
\newcommand{\pa}{\partial}

\newcommand{\Tr}{\mbox{Tr}}

\newcommand{\sgn}{\mbox{sgn}}

\renewcommand{\Im}{\mbox{Im}}

\newcommand{\+}{{\,+ \,}}

\def\dbend{\lower3.5pt\hbox{\manual\char127}}
\def\IL{\relax{\rm I\kern-.18em L}}
\def\IH{\relax{\rm I\kern-.18em H}}
\def\rlx{\relax\leavevmode}

\def\ZZ{\rlx\leavevmode\ifmmode\mathchoice{\hbox{\cmss Z\kern-.4em Z}}
 {\hbox{\cmss Z\kern-.4em Z}}{\lower.9pt\hbox{\cmsss Z\kern-.36em Z}}
 {\lower1.2pt\hbox{\cmsss Z\kern-.36em Z}}\else{\cmss Z\kern-.4em
 Z}\fi}

\newcommand{\erf}{ \textrm{erf}}

\title{Three Avatars of Mock Modularity}

\author{Atish Dabholkar and}
\author{Pavel Putrov}

\affiliation{International Centre for Theoretical Physics\\
Strada Costiera 11, Trieste 34151 Italy}

\abstract{
Mock theta functions were introduced by Ramanujan in 1920 but a proper understanding of mock modularity  has emerged   only recently with the work of Zwegers in 2002. In these lectures we  describe three manifestations of this apparently exotic mathematics  in  three important  physical contexts of holography, topology and duality where mock modularity has come to play an important role. 
 
\leftline{}
\leftline{}
\leftline{}
\leftline{\textsl{Lectures delivered at 2020 ICTP Online Summer School on String Theory and Related Topics}}}
\begin{document}
\maketitle
%\setcounter{tocdepth}{2}
%\tableofcontents

\section{Introduction}

In Ramanujan’s famous last letter to Hardy in 1920, he gave 17 examples (without any definition) of  mock theta functions that he found very interesting \cite{zbMATH00193455}. For example, 
\be
f(\tau) = -q^{-25/168} \sum_{n=1}^\infty \frac{q^{n^2}}{(1-q^n) \ldots (1 - q^{2n-1})} \, \qquad ( q:= e^{2\pi i \t} ) \, .
\ee
This function may seem unremarkable to the uninitiated, but Ramanujan had reasons to  believe that it was very similar to an ordinary theta function with hints of a `mock’ or `hidden’ symmetry under the modular group  $SL(2, \mathbb{Z})$. 

Despite much work by many eminent mathematicians, this fascinating mock modular symmetry remained  mysterious for a century until the thesis of  Zwegers in 2002 \cite{Zwegers:2008zna,MR2605321}. 
As we will see, the essence of mock modularity  is an incompatibility between holomorphy and modularity. For example, the function $f(\t)$ above, which is holomorphic in $\t$,  has no obvious modular properties. However, it is possible to add to it a  non-holomorphic `correction’ term such that the sum is indeed modular but at the expense of being nonholomorphic. One can either have modularity or holomorphy but not both.

It is useful to keep a geometric analogy in mind.  The property of being modular can be likened to being circular.  A  modular form is a function with  high degree of symmetry under the modular group $SL(2, \mathbb{Z})$ much like a  circle which is a geometric figure with a high degree of symmetry under the rotation group $O(2)$. Similarly, the property of being holomorphic can be likened to the property of being blue. In this analogy, a holomorphic mock modular form is  like a blue geometric figure which is not  circular but can be made circular by adding a non-blue piece to it.  One can either have circularity or blueness but not both.

Even though mock modular forms have a rich and interesting mathematical history, it was not clear if this exotic mathematics has any relevance to physics. A marvelous essay titled `A walk through Ramanujan’s Garden’ by Freeman Dyson \cite{Dyson:1987} on the occasion of the Ramanujan Centenary Conference in 1987 gives a glimpse of the history of these intriguing functions and ends with a prescient remark about possible physics applications: 

\textit{``My dream is that I will live to see the day when our young physicists, struggling 
to bring the predictions of superstring theory into correspondence with the facts of nature, 
will be led to enlarge their analytic machinery to include not only theta-functions but mock 
theta-functions\,\dots\; But before this can happen, the purely mathematical exploration of the mock-modular forms and their mock-symmetries must be carried a great deal further.’’}

True to the hope expressed by Dyson, over the past decade, mock modular forms have made their appearance  in diverse physical contexts connecting to deep and important physical concepts such as holography  and duality even if not directly to facts of nature.  The purpose of these lectures is to introduce the basic notions about mock modular forms and then outline their applications in three physical contexts: supersymmetric sigma models with non-compact target, counting of black hole degeneracies, and 4-dimensional Vafa-Witten topological gauge theory. We leave uncovered some other topics where mock modularity plays an important role, in particular, umbral moonshine \cite{Cheng:2012tq} and quantum 3-manifold invariants \cite{zbMATH01531028,Gukov:2016gkn,Cheng:2018vpl}. For other reviews of applications of (mock) modularity see for example \cite{zbMATH05808162,zbMATH05718024}.

\section{Modular Forms}
\label{sec:mockandmodular}

We first review modular and Jacobi forms and their appearance  in the context of black hole physics and holography, topology, and duality. For more details see \cite{Dabholkar:2012zz}

%\subsection{Modular Forms}
\label{sec:modularforms}

\subsection{Definitions}
Let $\IH$ be the upper half plane, \textit{i.e.}, the set of complex numbers $\tau$
whose imaginary part satisfies $\Im(\t)>0$. Let $SL(2, \mathbb{Z})$ be the group of matrices  $\begin{pmatrix} a & b \\ c & d \\  \end{pmatrix}$
 with integer entries such that $ad - bc =1$.

A \emph{modular form} $f (\t)$ of weight $k$ on $SL(2,\Z)$ is a holomorphic
function on $\IH$ that transforms as
\be
(c\t + d) ^{-k} f(\frac{a\tau + b}{c\tau + d})  = f(\tau) \qquad \forall \
\begin{pmatrix} a & b \\ c & d \\  \end{pmatrix}  \in SL(2, \mathbb{Z}) \, 
\ee
for an integer $k$ (necessarily even if $f \not \equiv 0$) and is bounded as $\rm{Im}(\t) \to \infty$. 
It follows from the definition that $f(\t)$ is periodic under $\tau \to \tau +1$ and can be written as a Fourier series
  \be\label{holmod} 
  f(\t) \ = \ \sum_{n= 0}^\infty a(n)\,q^n   \qquad \bigl(q :=  e^{2 \pi i \tau}\bigr)\,.  
  \ee
A holomorphic modular form must have non-negative weight (strictly positive weight if the form is not a constant). 

%\subsection{Types of Modular Forms}

If $a(0) =0$, then the modular form vanishes at infinity and is called a \emph{cusp form}.
In the other direction, one may weaken the growth condition as $\rm{Im}(\t) \to \infty$  from $f(\t) = \CO(1)$
to~$f(\t) = \CO(q^{-N})$  for some $N \ge 0$; then $f$ has a Fourier expansion 
containing finitely many negative powers of~$q$. 
%the Fourier coefficients of $f$ have the behavior $a(n)=0$ for $n < -N$. 
Such a function is called a \emph{weakly holomorphic modular form}.
The weight of weakly holomorphic modular forms can be negative. 

We denote the vector space over $\mathbb{C}$ of holomorphic modular forms of weight $k$ on~$SL(2,\IZ)$
by~$M_k$, while the space of cusp forms of weight $k$ and
the space of weakly holomorphic modular forms of weight $k$ are denoted by $S_k$
and $M^{\,!}_k$ respectively. We thus have the inclusions
  \be\label{in}  S_k \; \subseteq M_k \; \subset \;M^{\,!}_k  \, ,  \ee
The growth condition at $\rm{Im}(\t) \to \infty$ is related to the growth properties of the Fourier coefficients $a(n)$  as $n \to \infty$:

\bea \label{growth}
&& f \in S_{k}\;\Rightarrow \; a_{n} \= \CO(n^{k/2}) \quad \text{as} \quad n \to \infty\,; \\
&&  f \in M_{k}\; \Rightarrow \; a_{n} \= \CO(n^{k-1}) \quad \text{as} \quad n \to \infty\,; \\
&& f \in M_{k}^{\,!}\;\Rightarrow\; a_{n} \= \CO(e^{C \sqrt{n}}) \quad \text{as} \quad  n \to \infty 
\eea
The exponential growth of $a(n)$ for $f \in M_{k}^{\,!}$ can be recognized as the Cardy formula in conformal field theory as we explain below in an example.

\subsection{The Ring of Modular Forms}

%= 1 - 24*x - 72*x^2 - 96*x^3 - 168*x^4 - 144*x^5
 
 It is clear from the definition that the product  $ f_{1} (\t) f_{2} (\t)$ of two modular forms
 $f_{1}(\t)$ and  $f_{2}(\t)$ of weights $k_{1}$ and $k_{2}$ respectively is also a modular form of weight $k_{1} + k_{2}$. With these operation of usual pointwise addition and multiplication,  modular forms form a ring. 
An important theorem in the theory of modular forms states that the ring of (holomorphic) modular forms is generated by the two Eisenstein series $E_{4}(\t)$ and $E_{6}(\t)$ defined by
% Two  important modular forms on $SL(2,\Z)$ are the {\it Eisenstein series} $E_{4}$ and $E_{6}$ 
%$\in M_{k}$ ($k \ge 4$).  
\begin{eqnarray}\label{eisen}
   E_4 (\t) &\ = \ & 1 \+ 240\,\sum_{n=1}^\infty \frac{n^3q^n}{1- q^n} \ = \ 1+ 240 q + 2160q^2 + \cdots \, , \\
   E_6 (\t) &\ = & 1 \,-\, 504\,\sum_{n=1}^\infty \frac{n^5q^n}{1- q^n} \ = \ 1 -504q -16632q^2 - \cdots \,.
\end{eqnarray}
In other words,  any holomorphic modular form can be expressed as a linear combination of products of  $E_{4}$ and $E_{6}$  of appropriate weight. 

This important theorem reveals  the power of modularity. 
A priori, to determine an arbitrary translation invariant function, one needs to know all its infinite Fourier coefficients $\{ a(n) \}$ in the $q$-expansion \eqref{holmod}. Thanks to the theorem stated above, for a modular form, it suffices to know just the first few Fourier coefficients which can be determined  by comparison with \eqref{eisen}. 
For example, a weight $12$ modular form must be a linear combination of the form 
\be \label{weight12}
f_{12} (\t) = a E_{4}^{3} (\tau) + b E_{6}^{2} (\tau) \, ,
\ee
for some numerical complex coefficients $a$ and $b$ which can be readily determined knowing the first two Fourier coefficients. 

\begin{exercise}
 \label{ex:cusp}
 
Show that (up to normalization) there is a unique cusp form of weight $12$ and express it in terms of $E_{4}$ and $E_{6}$. This is an important example of a cusp form known as the discriminant function and denoted by $\D(\tau)$.

\end{exercise} 
%\subsection{Examples}

The  cusp form $\D(\tau)$ admits a product representation:
\be\label{discri}
\D(\tau) := q \prod_{n=1}^{\infty} (1 - q^{n})^{24} \,  = q \, - \, 24 q^2 \ + \  252 q^3 \ + \ \cdots 
\ee
The Fourier coefficients of the discriminant function, usually denoted by $\tau(n)$, play an important in number theory. Because $\Delta(\t)$ is a cusp form, $\t(n)$   grow very slowly as $n\rightarrow \infty$. 

In physics applications, one encounters not the cusp form but rather its inverse
\be\label{het}
\frac{1}{\D(\tau)} = \frac{1}{\eta(\tau)^{24}} := \sum_{n=-1}^{\infty} d(n) q^{n}
\ee
where 
\be
\eta(\tau) = q^{\frac{1}{24}} \prod_{n=1}^{\infty}(1- q^{n})
\ee
is the Dedekind $\eta$ function. This inverse of the cusp form is clearly a weakly holomorphic modular form, and arises, for example, as the partition function of $24$ left-moving bosons of the light-cone bosonic string. In number theory, the partition function above is well-known in the context of
the problem of partitions of integers. One can  identify
  \be\label{colorpartition}  d(n) \= p_{24}(n +1)  \qquad (n \geq 0)\,.  \ee
where $p_{24}(I)$ is the number of colored partitions of a positive integer $I$ using
integers of $24$ different colors, which is the same combinatoric problem of dividing energy $I$ among $24$  transverse oscillators.  Now, the Fourier coefficients grow exponentially  in accordance with \eqref{growth}
\be
d(n) \sim \exp [ 4\pi \sqrt{n} ] \, ,
\ee
and is related to the Cardy formula
\be
d(n) \sim \exp \left[ 2\pi\sqrt{\frac{ n c}{6}}\right] \, 
\ee
with $c = 24$ which the central charge of $24$ left-moving bosons.

\section{Jacobi Forms}
\label{sec:jacobiforms}

Jacobi forms are functions of two variables $\t$ and $z$ that transform nicely under the Jacobi transformations as defined below. One can regard $\t$ as the complex-structure parameter of a 2-torus obtained by modding a complex plane with coordinate $z$ by the lattice of points $z = \l \t + \m$ for all $ \l,\,\m \in \mathbb{Z}$. The Jacobi group is the group of Jacobi transformations as below generated by the modular transformations as well as the translations of $z$ by lattice shifts. 

\subsection{Definitions}

Consider a holomorphic function $\varphi(\tau, z)$ from $\IH \times \mathcal{C}$ to $\mathcal{C}$ which 
is ``modular in $\tau$ and elliptic in $z $’’ in the sense that it transforms under the modular group as
  \be\label{modtransform} 
   \vf\Bigl(\frac{a\t+b}{c\t+d},\frac{z}{c\t+d}\Bigr) \ =  \ 
   (c\t+d)^k\,e^{\frac{2\pi imc z^2}{c\t+d}}\,\vf(\t,z)  \qquad \forall \quad
   \Bigl(\begin{array}{cc} a&b\\ c&d \end{array} \Bigr) \in SL(2, \mathbb{Z}) \ee
and under the translations of $z$ by $\mathbb{Z} \tau + \mathbb{Z}$ as
  \be\label{elliptic}  \vf(\t, z+\l\tau+\mu)\ = \ e^{-2\pi i m(\l^2 \t + 2 \l z)} \vf(\t, z)
  \qquad \forall \quad \l,\,\m \in \mathbb{Z} \, , \ee
where $k$ is an integer and $m$ is a positive integer.

These equations include the periodicities $\vf(\t+1,z) = \vf(\t,z)$ and $\vf(\t,z+1) = \vf(\t,z)$, so $\vf$ has a Fourier expansion
  \be\label{fourierjacobi} \vf(\t,z) \ = \ \sum_{n, r} c(n, r)\,q^n\,y^r\,, \qquad\qquad
   (q :=e^{2\pi i \t}, \; y := e^{2 \pi i z}) \ . \ee
Equation \eqref{elliptic} is then equivalent to the periodicity property
%  \be\label{cnrprop}  c(n, r) \ = \ C(4 n m - r^2 ; \, r) \ ,
%  \qquad \mbox{where} \; C(\D ;r) \; \mbox{depends only on} \; r \, (\mod\, 2m) \ . \ee
%(When we want to emphasize the function $\vf$, we will denote $c(n,r)$ and $C(\D,r)$ by $c_\vf(n,r)$ and $C_\vf(\D,r)$.)
  \be\label{cnrprop}  c(n, r) \ = \ C(4 n m - r^2 , \, r) \ ,
  \qquad \mbox{where} \; C(\D, r) \; \mbox{depends only on} \; r \mod 2m \ . \ee

The function $\vf(\tau, z)$ is called a \emph{holomorphic Jacobi form} (or simply a \emph{Jacobi form})
of weight $k$ and index $m$ if the coefficients $C(\D,r)$ vanish for $\D<0$, {\it i.e.} if 
  \be\label{holjacobi}  c(n, r) \ = \ 0 \qquad \textrm{unless} \qquad 4mn \ge r^2\,. \ee
It is called a \emph{Jacobi cusp form} if it satisfies the stronger condition that
$C(\D,r)$ vanishes unless $\D$ is strictly positive, {\it i.e.}
  \be\label{cuspjacobi}  c(n, r) = 0 \qquad \textrm{unless} \qquad 4mn > r^2 \ , \ee
and it is called a \emph{weak Jacobi form} if it satisfies the weaker condition
  \be\label{weakjacobi} c(n, r) \ = \ 0\qquad   \textrm{unless}  \qquad n \geq 0 \, \ee
rather than \eqref{holjacobi}, whereas a merely \emph{weakly holomorphic Jacobi form} satisfies only the yet weaker condition
that $c(n,r)=0$ unless $n\ge n_0$ for some possibly negative integer~$n_0$ (or equivalently $C(\D,r)=0$ unless $\D\ge\D_0$
for some possibly negative integer~$\D_0$).  
%The space of all holomorphic (resp.~cuspidal, weak, or weakly holomorphic) Jacobi
%forms of weight~$k$ and index~$m$ will be denoted by $J_{k,m}$ (resp.~$J_{k,m}^0$,  $\wt J_{k,m}$, or $\wt J^{\,!}_{k,m}$).

Finally, the quantity $\D=4mn-r^2$, which by virtue of the above discussion is the crucial invariant of a monomial
$q^ny^r$ occurring in the Fourier expansion of~$\vf$, will be referred to as its {\it discriminant} (not to be confused with the discriminant function \eqref{discri} introduced earlier).  

\subsection{Theta Expansion}
\label{sec:theta}

If $\vf(\t, z)$ is a Jacobi form, then the transformation property (\ref{elliptic}) implies its
Fourier expansion with respect to $z$ has the form
  \be\label{jacobi-Fourier} \vf(\t, z) \= \sum_{\ell\inn \Z} \;q^{\ell^2/4m}\;h_\ell(\t) \; e^{2\pi i\ell z} \ee
where $h_\ell(\tau)$ is periodic in $\ell$ with period $2m$.  In terms of the coefficients \eqref{cnrprop} we have
  \be\label{defhltau}  h_{\ell}(\t) \= \sum_{\D} C(\D,\ell) \,  q^{\D/4m} \, \qquad \qquad (\ell \inn \Z/2m \Z)\;.  \ee
Because of the periodicity property, equation \eqref{jacobi-Fourier} can be rewritten in the form 
  \be\label{jacobi-theta} \vf(\t,z) = \sum_{\ell\inn \Z/2m\Z} h_\ell(\t) \, \vth_{m,\ell}(\t, z)\,, \ee
where $\vth_{m,\ell}(\t,z)$ denotes the standard index $m$ theta function 
\be
\vartheta_{m,\ell}(\tau, z) 
 := \sum_{n\in \mathbb{Z}} \,q^{(\ell+2mn)^2/4m} \, y^{\ell+2mn}
 \ee
(which is a Jacobi form of weight $\half$ and index $m$ on some subgroup of
$SL(2, \mathbb{Z})$). This is called the theta expansion of $\vf$.  The coefficiens $h_{\ell}(\t)$ are  modular forms of weight $k-\frac12$
and are weakly holomorphic, holomorphic or cuspidal if $\vf$ is a weak Jacobi form, a Jacobi form or a Jacobi cusp form, respectively. 
More precisely, the vector $h := ( h_1, \ldots, h_{2m})$ transforms like a modular form of weight $k-\frac 12$ under $SL(2,\Z)$.  Jacobi forms, which are functions are two complex variables $\t$ and $z$  are thus equivalent to vector valued modular forms of a single complex variable $\t$. 

\subsection{The Ring of Jacobi Forms}

Consider the two Jacobi forms defined below
\bea
 A(\tau, z) =&  \frac{\vartheta_1^2( \tau, z)}{\eta^6(\tau)}  \qquad \qquad  & (k=-2 \, , \, m=1) \label{A} \\
B(\tau, z) = & 8 \left(\frac{\vartheta_2^2( \tau, z)}{\vartheta_2^2( \tau)} + \frac{\vartheta_2^2( \tau, z)}{\vartheta_2^2( \tau)}  + \frac{\vartheta_2^2( \tau, z)}{\vartheta_2^2( \tau)}  \right)  \qquad \qquad  & (k= 0 \, , \, m=1) \label{B}
\eea
where $\{\vartheta_{i}(\t, z)\} $ are the Jacobi theta functions. The ring of Jacobi forms of weight $k$ and index $m$ is generated by these two Jacobi forms with ordinary modular forms as coefficients. 
Index and weight both add when you multiply Jacobi forms and modular forms. Thus, as in the case of modular forms, a Jacobi form can be determined by knowing its weight and index and first few Fourier coefficients. 
For example, the most general holomorphic Jacobi form of weight $4$ and index $2$ must be of the form
\be
a \, E_4^2(\tau) \, A^2 (\tau, z) +  \, b  E_6(\tau) \, A (\tau, z) \, B (\tau, z)\, + c\,E_4(\tau) \, B^2(\tau, z)
\ee
and hence is completely fixed by determining the three constants $a, b, c$. 

\section{Modularity in Physics}

The modular group $SL(2, \mathbb{Z})$ appears in physics in many different contexts as a physical symmetry.
The most familiar occurrence  is  in conformal field theory. The modular group of a two-torus is the group of global diffeomorphisms of the 2-torus modulo the Weyl group. If one uses a coordinate invariant regulator then the path integral for the sigma model on a two-dimensional torus is diffeomorphism invariant. For a conformal sigma model  it also Weyl invariant. Hence the path integral of a conformal field theory on a 2-torus is expected to be  $SL(2, \mathbb{Z}) $-invariant.   It is thus natural that modular forms, which transform nicely under this symmetry, have come to play an important role in physics. We focus on the following three contexts.

\subsection{Topology}

In general, the partition function of a conformal field theory will be a function of both $\t$ and $\bar \t$. However, certain indexed partition functions, which capture a topological subsector of the theory are  functions of $\t$ alone and are thus holomorphic by an argument due to Witten that we outline in \S\ref{witten}.

Consider a conformally invariant nonlinear sigma model with target space  $M$ with  $(2, 2)$ superconformal symmetry with left and right super Virasoro algebra and central charge $6m$ both for the left-movers and the right-movers. There is a $U(1)_{L} \times U(1)_{R}$ R-symmetry of the superconformal algebra. The $U(1)_{L}$ symmetry (and similarly the  $U(1)_{R}$ symmetry ) is anomalous with anomaly coefficient $m$ which by supersymmetry is related to the conformal anomaly governed by the central charge $6m$.  Let $F_{R}$ and $F_{L}$ be the charges associated with this symmetry.  One can then define  the elliptic genus as the following indexed  partition function:
\be
\label{elliptic-def}
\chi (\tau, z| M) =\mathrm{Tr} \,  (-1)^{ F_R + F_L} \, {e}^{2 \pi i \tau H_L}\,  e^{-2 \pi i \bar{\tau} H_R} \, e^{2 \pi i z F_L} \, .
\ee

Elliptic genus thus defined is a Jacobi form weight $0$ and index $m$.  As argued above, it is modular invariant and hence has  modular weight zero.  The $(2, 2)$ superconformal algebra has an additional symmetry called the spectral flow symmetry. If one bosonizes the left-moving $U(1)_{L}$ current, then the spectral flow symmetry corresponds to the shifts of this boson. Consequently,   the path integral is elliptic with index $m$ related to the anomaly in the $U(1)_{L}$ symmetry. 
Because, the R-symmetry acts on fermions alone, $(-1)^{ F_R}$ can be identified with the right-moving fermion number. The indexed partition function defined above can thus be thought of as the Witten index  for the right-moving sector that counts right-moving ground states which is a topological quantity called the elliptic genus. 

The Fourier coefficients of the elliptic genus correspond the Dirac indices of an infinity family of Dirac-like (elliptic) operators.  Jacobi forms thus appear naturally in topology in the context of elliptic genera of manifolds.

\begin{exercise}
 \label{ex:T4genus}
 
Show that the modified elliptic genus of $T^{4}$ equals $A(\tau, z)$.
\end{exercise} 

\begin{exercise}
 \label{ex:K3genus}
 
Show that the elliptic genus of $K3$ equals $B(\tau, z)$.
\end{exercise}

\subsection{Duality}

The hypothesis of  $S$-duality asserts that
 $\scN=4$ super Yang-Mills theory is invariant under the action of a large duality group ($SL(2,\mathbb{Z})$ or a close relative, depending on
the four-dimensional gauge group $G$) acting on $\tau\equiv \tau_1+i\tau_2=\theta/2\pi+4\pi i/g^2$; 
here $g$ and $\theta$ are the gauge coupling and theta angle; $\tau_1$ and $\tau_2$ denote the real and imaginary parts of $\tau$.
But $S$-duality is hard to test, because  computations for strong coupling are difficult.   One way to circumvent this difficulty
is to consider a topologically twisted version of the theory in which localization can be used to perform computations for strong coupling.

A twist of particular interest is the Vafa-Witten twist introduced in \cite{Vafa:1994tf}.   With this twisting, a formal argument shows that the partition function on a compact four-manifold $M^4$ is holomorphic in $\tau$
or equivalently in $q=\exp(2\pi i\tau)$.  Furthermore, if a certain curvature condition (eqn. (2.58) in  \cite{Vafa:1994tf}) is satisfied, the evaluation of the path integral can formally be
argued to
localize  on the contribution of ordinary Yang-Mills instantons.  The contribution to the path integral from the component of field space with instanton number\footnote{Here $n$ is an integer
for a simply-connected gauge group such as $G=SU(2)$, but may have a fractional part if  $G$ is not simply-connected.   The fractional part is  determined
by a two-dimensional cohomology class (for example, by the second Stieffel-Whitney class $w_2$ if $G=SO(3)$), and in the partition function  $\sum_n a_n q^n$, it is natural
to sum 
over all bundles  keeping this class fixed.   The values of $n$ in the sum are then congruent to each other mod  $\Z$.  A restriction on $w_2$ (and its analog for other groups)
is assumed in eqn. (\ref{expz}).}  $n$ is then $a_n q^n$,
where $a_n$ is the Euler characteristic of the instanton number $n$ moduli space $\scM_n$.   Thus the partition function after  summing over bundles
of all values of the instanton number is expected to be
\bea\label{expz} Z=\sum_n a_n q^n. \eea

The relevant curvature condition is  highly restrictive, but there are a number of four-manifolds that satisfy this condition
and for which computations of the $a_n$ were available in the mathematical literature  \cite{Klyachko:1991, Yoshioka:1994}.   
In particular, two important examples that we consider here are a $K3$ surface and $\CP^2$. 

%For a four-dimensional quantum gauge theory, a congruence subgroup of
%$SL(2, \mathbb{Z})$ is the S-duality group which in general maps the weak coupling regime of the theory to  the strong coupling regime. Vafa-Witten theory is a twisted version of the gauge theory on a Euclidean 4-manifold $\bX$ which focuses on certain topological sector  of the theory. Observables in this sector, and in particular the partition function of the twisted gauge theory, is often computable using localization methods even at strong coupling, thus contains nontrivial information about the nonperturbative structure of the theory. 

A four-dimensional  gauge theory arises naturally as the low-energy worldvolume theory of $N$ conincident  $D3$-branes. In the Euclidean version one can consider Euclidean $D3$-branes wrapping a four-manifold $M^4$. In the $D$-brane picture, instantons in the  gauge theory correspond to $D$-instantons bound to the $D3$-branes. The Vafa-Witten partition function counts the number of such bound states. By T-duality in directions transverse to the branes,  this is the same the number of $D0$-$D4$ or $D1$-$D5$ bound states which we will encounter in the context of black hole microstates. 

In the simplest context of a single Euclidean $D3$ brane wrapping a $K3$ with $m+1$ point-like instantons\footnote{In a $U(1)$ gauge theory the instantons are singular but one can turn on a noncommmutativy parameter to make the problem better defined},   one can heuristically think of them as  $m+1$ point particles moving on $K3$.  The indexed partition function for each particle gives simply the ground states of the supersymmetric quantum mechanics of a super particle with $K3$ as the target, which is nothing but Euler characteristic of $K3$ which is $24$.  The $(m+1)$ superparticles on $K3$ can be treated as identical bosonic particles and thus have as the target space symmetrized product of $(m+1)$ copies of $K3$.  The indexed partition function which gives the coefficients $a_{m+1}$   thus equal the orbifold Euler character $ \chi(\textrm{Sym}^{m+1}(K3))$ of the
symmetric product of $(m+1)$ copies of $K3$-surface \cite{Vafa:1994tf}. 
The
generating function for the orbifold Euler character 
 \be\label{Zeuler} Z(\s) = \sum_{m=-1}^\infty \chi(\textrm{Sym}^{m+1}(K3))\, p^m
  \qquad \bigl(p:= e^{2\pi i\s}\bigr)\ee
can be thought of  the grand-canonical partition function for this quantum mechanical system of $m+1$ identical bosons with fugacity $p$. This can be readily evaluated using Bose-Einstein distribution knowing the single-particle degenercies  to obtain
%\cite{Goettsche:1990go} 
to obtain
  \be\label{Zeuler2}  Z(\sigma) =
  \frac{1}{p} \prod_{n=1}^{\infty} \frac {1}{(1 - p^{n})^{24}} \,. \ee
  This is a modular form of weight $-12$, in fact the inverse of the cusp form that we have encountered earlier.  Its modular properties are a consequence of S-duality of the gauge theory living on the D-brane worldvolume. 
  
Note that this partition function is the same as the one we encountered for $24$ left-moving bosons of the heterotic string. This is not an accident. The heterotic string on a 4-torus $T^{4}$ is dual to  Type-IIA string on $K3$
Duality requires that the number of  BPS-states of a given charge must equal the
number of BPS-states with the dual charge. The equality of the two partition functions
(\ref{het}) and (\ref{Zeuler2}) coming from two very different counting
problems is consistent with this expectation. This fact was indeed one of the
early indications of a possible duality between heterotic and Type-II strings
\cite{Vafa:1994tf}.

%
%The DH-states correspond to the microstates of a small black hole 
%\cite{Sen:1995in,Dabholkar:2004yr,Dabholkar:2004dq} for large $n$. The macroscopic entropy
%$S(n)$ of these black holes should equal the asymptotic growth of the degeneracy
%by the Boltzmann relation \be\label{bolt} S(n) = \log d(n); \quad n
%\gg 1 \,. \ee 
%
%In the present context, the macroscopic entropy can be
%evaluated from the supergravity solution of small black holes 
%\cite{LopesCardoso:1998wt,LopesCardoso:1999ur,LopesCardoso:1999cv,
%LopesCardoso:1999xn,Dabholkar:2004yr,Dabholkar:2004dq}. 
%The asymptotic growth of the microscopic degeneracy can
%be evaluated using the Hardy-Ramanujan expansion (Cardy formula). There is a
%beautiful agreement between the two results 
%\cite{Dabholkar:2004yr,Kraus:2005vz}
%  \be\label{bolt2} S(n) \= \log d(n) \sim 4 \pi \sqrt{n}\quad n \gg 1 \,.   \ee
%Given the growth properties of the Fourier coefficients mentioned above, it is clear
%that, for a black hole whose entropy scales as a power of $n$ and not as
%$\log(n)$, the partition function counting its microstates can be only weakly
%holomorphic and not holomorphic.

\subsection{Holography}

In $AdS_{3}$ quantum gravity, the modular group  $SL(2, \mathbb{Z})$ is the group of global diffeomorphisms of the boundary. Quantum string theory in $AdS_{3}$  background is expected to be dual to a sigma model living on the boundary.

A well-studied special case of this duality is in the context of  Type-IIB string theory compactified on $K3 \times S^{1}$. Consider $Q_{5}$ $D5$-branes wrapping $K3 \times S^{1}$ and $Q_{1}$ $D1$-branes wrapping $S^{1}$ with $Q_1 Q_5 \equiv m $. The near horizon geometry of this brane system is $AdS_{3} \times S_{3} \times K3$. 
The sigma model dual to this theory has as target space  symmetric product of $m+1$ copies of $K3$ defines a $(2, 2)$ superconformal symmetry (actually it has a larger $(4, 4)$ symmetry but that is not relevant to our discussion) with left and right moving central charges $6(m+1)$.  The elliptic genus defined as above
\be
\chi (\t,z | \mathrm{Sym}^{m+1}\,K3) \, ,
\ee
gives the indexed partition function of this boundary theory is a generalization of the Euler characteristic of symmetrized product that we encountered earlier. By general arguments reviewed earlier, this elliptic genus is  a weak Jacobi form of weight $0$ and index $m+1$. Henceforth we will denote it simply by $\chi_{m+1} (\t,z )$. We thus see that Jacobi forms appear naturally in the context of $AdS_{3}/CFT_{2}$ holographic duality. 

\section{Mock Modular Forms}

We now review mock modular  and mock Jacobi forms. 

\label{sec:mockmodularforms}

\subsection{Definitions}

We define a (weakly holomorphic) {\it pure mock modular form} of weight $k\in \frac12 \Z$ as the first member 
of a pair $(h,g)$, where
\begin{enumerate}
\item $h$ is a holomorphic function in $\IH$ with at most exponential growth at all cusps, 
\item the function $g(\t)$, called the \textit{shadow} of $f$, is a holomorphic\footnote{One can also consider the case where the shadow is allowed
to be a weakly holomorphic modular form, but we do not do this since none of our examples will be of this type.} modular form of weight $2-k\,$, 
and 
\item the sum $\wh h := h\+g^*$, called the {\it completion} of $f$, transforms like a holomorphic modular form of weight $k$,
   {\it i.e.} $\wh h (\tau)/\theta (\tau)^{2k}$ is invariant under $\t \to \g \t$  for all $ \t \in \IH $ and for all
   $\g$ in some congruence subgroup of $SL(2,\Z)$. 
\end{enumerate} 
Here $g^*(\t)$, called the {\it non-holomorphic Eichler integral}, is a solution of the differential equation
\be\label{starinv} (4\pi\t_2)^k\,\frac{\pa g^*(\t)}{\pa \bar{\tau}} \= -2\pi i\;\overline{g(\tau)} \, . \ee
If $g$ has the Fourier expansion $g(\t)=\sum_{n\ge0}\;b_n\,q^n$, we fix the choice of $g^*$ by setting 
\be\label{defstar} 
  g^*(\t)  \=  \bar b_0\,\frac{(4\pi\t_2)^{-k+1}}{k-1} \+ \sum_{n>0} \;n^{k-1}\,\bar b_n \;\G(1-k,4\pi n\t_2)\;q^{-n}\, , 
\ee
where $\t_2 = \rm{Im}(\t)$ and 
\be\label{incomp}
\G(1-k,x) = \int_x^{\infty} t^{-k}\,e^{-t}\,dt
\ee
 denotes the incomplete gamma function.
% and where 
%the first term must be replaced by $ -\bar b_0\,\log(4\pi\t_2)\,$ if $k=1$. 

Note that the series in~\eqref{defstar} converges despite 
the exponentially large factor $q^{-n}$ because $\G(1-k,x)= O(x^{-k}e^{-x})\,$. If we assume either that $k>1$ or that $b_{0}=0$,
then we can define $g^*$ alternatively by the integral  
\be\label{Lkintrep} g^*(\t) = \biggl(\frac i{2\pi}\Bigr)^{k-1} \int_{-\bar{\t}}^{\infty} (z+ \t)^{-k} \ \overline{g(-\bar{z})}\; dz \;. \ee
(The integral is independent of the path chosen because the integrand is holomorphic in $z$.) 
Since $h$ is holomorphic, \eqref{starinv} implies that the completion of $h$ is related to its shadow by
% A basic property of the completion, immediate from \eqref{starinv} because $h$ is holomorphic, is that 
% Because $h$ is holomorphic, \eqref{starinv} implies the key property 
    \be\label{ddtbarh}   (4\pi\t_2)^k\,\,\frac{\pa \wh h(\t)}{\pa \bar{\tau}} \= -2\pi i\;\overline{g(\tau)}\;.  \ee
% of the completion. 
This `\textit{holomorphic anomaly equation}' captures the failure of holomorphy of the completion of the mock modular form,  which is modular but not holomorphic.

If one multiplies the mock modular form by a holomorphic modular form $f(\t)$  of weight $k'$ the holomorphic anomaly equation above simply gets multiplied by $f(\tau)$ on both sides.  One thus gets a \textit{mixed} mock modular form $f(\t) h(\t)$ of weight $k+ k'$ with completion $f(\t)\hat h(\t)$ and a mixed  shadow $\bar f(\tau) g(\tau)$ that is a product of a holomorphic and anti-holomorphic piece.

 \subsection{Examples}
 
Apart from the examples introduced by Ramanujan,  a particularly interesting example is the Zagier mock modular form which will play a role  in our later discussions. The Fourier coefficients of this mock modular form give the
Hurwitz-Kronecker class numbers, denoted by  $H(N)$ for $N \in \bZ$.  

These class numbers  are defined for $N>0$ as the number of $PSL(2,\bZ)$-equivalence classes of 
integral binary quadratic forms of discriminant $-N$, weighted by the reciprocal of the number of 
their automorphisms (if $-N$ is the discriminant of an imaginary quadratic field $K$ other than $\bQ(i)$ or 
$\bQ(\sqrt{-3})$, this is just the class number of $K$), and for other values of $N$ by 
$H(0) = -1/12$ and $H(N)=0$ for $N<0$. These numbers vanish unless $N$ is $0$ or $-1$ modulo $4$. 

The Zagier mock modular form  \cite{Zagier:1975a} is  the generating  function for  these class numbers:
\be\label{classnum}  
{\bf H}(\t) \;:=\; \sum_{N=0}^{\infty} H(N)\,q^{N} 
\ = -\frac{1}{12} \ + \frac{1}{3} q^3 \ + \frac{1}{2} q^4 \+ q^7 \ + q^8 \ + q^{11} \+ \cdots  
\ee
It   is a `pure mock modular form’  
of weight $3/2$ on $\Gamma_0(4)$ and  shadow the classical theta function $\vartheta (\tau)  = \sum q^{n^2}$.
See  \cite{MR2605321,Dabholkar:2012nd} for the definition and further discussion.
We see from the $q$-expansion of ${\bf H}(\tau)$ that it   has no poles at $q=0$  and hence  is strongly holomorphic.  Consequently,  its Fourier coefficients grow very slowly. This  is  exceptional. In fact, up to minor variations the Zagier mock modular form is essentially the {\it only} known non-trivial example of a strongly holomorphic pure mock modular form.

\section{Mock Jacobi Forms}

A mock Jacobi form is a  new mathematical object defined in \cite{Dabholkar:2012nd} as a mock generalization of a Jacobi form. It has the same elliptic symmetry  properties as a Jacobi form and hence does admit a theta expansion using this symmetry.   However,  the theta-coefficients are not vector-valued  modular forms but rather vector-valued mock modular forms. 

\subsection{Definitions}

By a (pure) {\it mock Jacobi form}   (resp.~{\it weak mock Jacobi form}) of weight $k$ and index $m$  we will mean \cite{Dabholkar:2012nd} a 
holomorphic function $\vf$ on $\mathbb{H} \times \mathbb{C}$ that satisfies the elliptic transformation property~\eqref{elliptic},
and hence has a Fourier expansion as in~\eqref{fourierjacobi} with the periodicity property~\eqref{cnrprop}
and a theta expansion as in~\eqref{jacobi-theta}, and that also satisfies the same cusp conditions~\eqref{holjacobi}
(resp.~\eqref{weakjacobi}) as in the classical case, but in which the modularity property with respect to the
action of $SL(2,\Z)$  is weakened: the coefficients $h_{\ell}(\tau)$ in \eqref{jacobi-theta}
are now mock modular forms rather than modular forms of weight $k-\frac12$, and the modularity property of
$\vf$ is that the {\it completed} function 
  \be
  \label{defphihat} 
  \wh \vf(\t,z) \;=\!\sum_{\ell\inn\Z/2m\Z} \wh h_\ell(\t) \, \vth_{m,\ell}(\t,z)\,,  
  \ee
rather than $\vf$ itself, transforms according to \eqref{modtransform}.  If $g_\ell$ denotes the shadow of $h_\ell$, then we have
  $$ \wh\vf(\t,z) \;= \! \vf(\t,z)\+\sum_{\ell\inn\Z/2m\Z} g^*_\ell(\t)\,\vth_{m,\ell}(\t,z)$$
with $g^*_\ell$ as in \eqref{defstar} and hence, by \eqref{starinv},   
 \be 
 \psi(\t,z)\;:=\; \t_2^{k-1/2}\,\frac{\pa}{\pa\overline\t}\wh\vf(\t,z)\;\doteq\;\sum_{\ell\inn\Z/2m\Z} \overline{g_\ell(\t)}\,\vth_{m,\ell}(\t,z)\,. 
 \ee
(Here $\doteq$ indicates an omitted constant.)  
\label{sec:mockjacobiforms}

\subsection{Example}
\label{sec:mock-example}

Using the class numbers introduced earlier one can define
\bea\label{fs}
\hh_0(\t) &=& \sum_{n\geq 0}H(4n) q^{n}+
2{\tau_2}^{-1/2}\sum_{n\in Z} \beta (4\pi n^2\tau_2 ) q^{-n^2}\,, \nonumber \\
\hh_1(\t) &=&\sum_{n>0} H(4n-1)q^{n-\frac{1}{4}}+
2{\tau_2}^{-1/2}\sum_{n\in Z} \beta (4\pi (n+\frac{1}{2})^2
\tau_2 ) q^{-(n+\frac{1}{2})^2},
\eea
with $\tau = \t_{1} + i \t_{2}$ and 
\be 
\beta (t)={1\over 16 \pi}\int_1^\infty u^{-3/2} {\rm exp}(-u t)\ du \, ,
\ee
which equals  the complementary error function up to normalization and can be easily related to the incomplete Gamma function introduced earlier. 

The functions $\{\hh_{\ell}(\tau)\}$ are not purely holomorphic because of the second term in \eqref{fs}
and  satisfy the \textit{holomorphic anomaly equation}:
\bea\label{holano}
\tau_2^{3/2} {\partial\over \partial\bar \tau} \hh_0&=&{1\over 16 \pi i}
{\sum_{n\in
		\bZ}{\bar q}^{ n ^2}}\,,
		\label{hol-anomaly-vector}
		\\
\tau_2^{3/2}{\partial\over \partial\bar \tau} \hh_{1}&=&{1\over 16 \pi i}
{\sum_{n\in
		\bZ}{\bar q}^{ (n+{1\over 2})^2}}\,.
\label{hol-anomaly-vector-1}
\eea
A nontrivial fact \cite{Zagier:1975a} is that $\hh(\t)=\begin{pmatrix}\hh_0(\tau)\cr \hh_1(\tau)\end{pmatrix}$  transforms 
as a vector valued modular form with weight $3/2$ under the modular group $\Gamma_0(4)$. In particular, 
\bea
\left(\begin{array}{c}\hh_0(-1/\tau) \\
	\hh_1(-1/\tau)
\end{array}\right)
=\left({\tau\over i}
\right)^{3/2}\cdot {-1\over \sqrt 2}
\left(\begin{array}{cc}
	1 & 1 \\1 &-1\end{array}\right)
\left(\begin{array}{c}\hh_0(\tau) \\ \hh_1(\tau) 
\end{array}\right)      \, .             
\eea
Using the definitions introduced earlier, the holomorphic parts
%In modern terminology  \cite{Zwegers:2008zna,MR2605321,Dabholkar:2012nd}, 
\bea\label{hs}
h_0(\t) &=& \sum_{n=0}^\infty H(4n) q^n\,, \\ 
h_1(\t) &=&  \sum_{n=1}^\infty H(4n-1)q^{n-{1\over 4}}\,.
\eea
are components of  a \textit{ vector-valued pure  mock modular form}  $h(\t)$  with  a holomorphic \textit{shadow} $g(\t)$ with components
\bea\label{gs}
g_0(\t) &=& c
{\sum_{n\in
		\bZ}{ q}^{ n ^2}}\,,\\
g_{1}(\t) &=& c
{\sum_{n\in
		\bZ}{ q}^{ (n+{1\over 2})^2}}\, , 
\eea
where $c = {1\over 16 \pi i}$ is  the overall normalization for which there is no standard convention. 
The vector  $h(\t) $ is holomorphic but  not modular whereas the vector $\hat h(\t)$ is modular but not holomorphic. 
%
%Addition of the correction terms in \eqref{fs} constructed from the  shadow vector $g(\t)$  yields its  \textit{modular completion} $\hh (\t)$. 
%The completion is modular but not holomorphic. This incompatibility between modularity and holomorphy is the essence of mock modularity. 

The components $\{ h_{\ell}\}$ are most naturally regarded as the vector of theta coefficients of a mock Jacobi form $ \cH(\t,z)$ defined by
\be\label{classH}  
\cH(\t,z) \,  := \, h_0(\t)\vth_{1,0}(\t,z)\+ h_1(\t)\vth_{1,1}(\t,z) 
\, = \,  \sum_{\substack{n,\,r\in{\bZ} \\ 
		4 n - r^2 \ge 0}}  H(4n-r^2) \, q^n\,y^r\,,  
\ee 
%where $y=e^{2\pi i z}$ and
%\be\label{levelm}
%\vth_{m,\ell}(\t,z) :=  \sum_{\substack{{r\inn\Zint} \\ {r\,\equiv\,\ell\; (\textrm{mod}\,2m)}}} q^{r^2/4m} \, y^r \, \qquad \qquad (\ell\  \textrm{mod} \ 2m)
%\ee
%are the level $m$ theta functions. 
%
Following the definition in above, one can check that $ \cH(\t,z)$ 
is a mock Jacobi form of weight $2$ and index $1$ with holomorphic anomaly (up to  normalization)
\be
\overline{\vth_{1,0}(\t,0)}\,\vth_{1,0}(\t,z)\,  + \,
\overline{\vth_{1,1}(\t,0)}\,\vth_{1,1}(\t,z) \, .
\ee

%
%The connection to mock modularity can be seen more simply by combining the two components of the vector-valued mock modular form into  a single  mock modular form defined by ${\bf H}(\t)=h_0(4\tau)+h_1(4\tau)$. 

\section{Black Holes and Mock Modularity}
\label{sec:blackholes}
Consider  Type-IIB theory compactified on $K3 \times T^{2}$ which results in a four-dimensional theory with $\cN=4$ supersymmetry. A dyonic state is specified by the  with charge vector electric charge vector $Q$ and magnetic charge vector $P$. There are three  T-duality invariants\footnote{In addition, there is an arithmetic U-duality invariant given by
$I = \gcd (Q \wedge P)$
that is relevant to the counting problem. 
We assume henceforth  that $I=1$.} that  given by
\be
m = P^{2}/2 \, , \qquad n = Q^{2}/2 \, , \qquad \ell =  Q\cdot P \, .
\ee

\subsection{Black Holes, Poles, and Walls}

The  indexed degeneracies of such dyonic states are given by \cite{Dijkgraaf:1996it,Dijkgraaf:2002ac,Gaiotto:2005gf, David:2006yn} the Fourier
coefficients of a meromorphic Jacobi form $\psi_m(\t, z)$ of weight
$-10$ and index $m$:
\be\label{meroformtwo} 
\psi_m(\t, z) = \frac{\chi_{m+1} (\t,z)}{A(\t,z)} \,
\frac{1}{\Delta(\t)} \, \,  \,
, \ee 

This poses a puzzle. The counting function has a double pole at $z=0$:
\be
\frac{ p_{24}(m+1)}{ \Delta(\tau)}  \frac{1}{z^{2}}
\ee
and at all the lattice images $z = \l\t + \m$. 
Because of the factor $A(z, \t)$ in the
denominator, $\psi_{m}$ is a \textit{meromorphic} Jacobi form with a 
double pole at $z=0$ and its images under elliptic transformations, thus at all points 
$z = \l \t + \m$. This would mean that the answer for different contours is different and the degeneracies are not uniquely defined. How is this possible?

%\subsection{Poles and Walls}

A physical resolution of this puzzle involves what is known as the wall-crossing phenomenon.
The moduli space of the theory is divided into chambers separated by walls, or codimension one surfaces. The quantum degeneracies jump upon crossing walls and therefore indeed are not uniquely defined. This non-uniqueness is related precisely to the   ambiguity in choosing the contour to obtain the Fourier coefficients. 

One obtains a beautifully consistent picture. The choice of the contour $\cC(\m)$ depends on the moduli $\m$.  Crossing a wall in the moduli space corresponds to crossing a pole. Changing the moduli deforms the contour but as long as one does not encounter a pole, one remains in a  given  chamber in the moduli space and the degeneracies do not change. Upon crossing a pole, the degeneracies jump corresponding to the  jump upon crossing a wall. This jump in the degeneracies is given by the residue at the
pole encountered.

\subsection{Multi-Centered Black Holes and the Appel-Lerch Sum}

The wall-crossing phenomenon however raises another puzzle. One of the main reasons for studying black hole microstates is that these microstates is an intrinsic property of the black hole horizon. One believes that therefore one can learn something about the short-distance structure of the theory from studying these microstates. If this is the case, then
the quantum degeneracy of a black hole should not depend on asymptotic moduli. How come then we get a different answer depending on which region of the moduli space we are in?

The resolution of this puzzle involves multi-centered black hole solutions. 
For fixed value of the magnetic charge, the partition function
$\psi_{m}(\tau,z)$ contains all the information about the \textit{asymptotic} quantum degeneracies.  
The indexed degeneracies receive  contributions both from single centered black holes as well as the multi-centered black holes. The multi-centered black holes are in a bound state at  fixed distance from each other which is determined by the asymptotic moduli. As the moduli are varied, this distance varies. In particular, the distance goes to infinity as  a wall is approached, which means that on the other side of the wall these bound states no longer exit and hence these multi-centered black holes no longer contribute to the indexed degeneracies.  Thus, even though the degeneracy of a black hole is indeed a property of its horizon, the  degeneracies of multi-centered black
holes which are bound states can jump upon crossing walls in moduli space depending on whether a stable bound state exists or not. 

This poses an interesting question whether one can define a counting function whose Fourier coefficients give directly the degeneracies of the horizons of the single-centered black holes without the polluting contributions from the multi-centered black holes. After all, the degrees of freedom of the horizon is the quantity of real interest in the explorations of quantum gravity.
It turns out that one can in fact remove this ambiguity in Fourier expansion
by subtracting the contribution of the various multi-centered black holes which is captured by the function
\be\label{Tm} 
 \psi^{P}_m \; := \; \frac{p_{24}(m+1) }{\eta^{24}(\tau)} \, \sum_{s\in\Zint} \, \frac{q^{ms^2 +s}y^{2ms+1}}{(1 -q^s y)^2} \ , 
\ee
which is simply the `elliptic average’ over the lattice $ \mathbb{Z} \tau + \Zint$ of the simplest wall-crossing at $z =0$ given by 
\be
  \frac{p_{24}(m+1)}{\eta^{24}(\tau)} \frac{y}{(1-y)^2}\,.
  \ee
The function $\psi_{m}^{P}$ is called the {\it polar} part of $\psi_{m}$ because it has identical poles in the complex $z$ plane. It is convenient to define
\be\label{rescale1}
 \psi^{P}_{m}(\tau, z) \; := \; \frac{p_{24}(m+1) }{\eta^{24}(\tau)}    \cA_{2, m}(\tau, z)
 \ee
where $\cA_{2, m}(\tau, z)$ is known as the Appel-Lerch sum which is simply the elliptic average of $1/z^{2}$ defined by
\be
\cA_{2, m}(\tau, z) = \sum_{s\in\Zint} \, \frac{q^{ms^2 +s}y^{2ms+1}}{(1 -q^s y)^2} \ , 
\ee
One can now obtain the horizon degeneracies of single-centered black holes simply by subtracting this contribution from the partition function $\psi_{m} (\t, z)$ which counts all degeneracies. 
This physical observation corresponds to the following decomposition theorem which also connects this problem with mock modularity.

\subsection{The Decomposition Theorem}
\label{sec:decomposition}

The decomposition theorem \cite{Dabholkar:2012nd} states that subtracting $\psi^{P}_{m}$ leaves us with the finite part 
$\psi_{m}^{F}$, which, as we saw above is exactly the attractor partition function:
\be\label{singlepart}  
%  Z^*_{m} (\tau,z) \ = \ 
  \psi_{m}^{F}(\tau,z) \; \equiv \; \psi_{m}(\tau,z) - \psi^{P}_{m}(\tau,z) \ .  
\ee
By separating part of the function $\psi_{m}$, we have of course broken modularity, and $\psi_{m}^{F}(\t,z)$  
is not a Jacobi form any more.
However, $\psi_{m}^{F}$ still has a very special modular behavior, in that 
it  is a mock Jacobi form. It is convenient to define as in \eqref{rescale1} a function 
$\widehat{\vf_{m}^{F}}(\tau)$ by 
\be\label{rescale2}
 \psi^{F}_{m}(\tau, z) \; := \; \frac{p_{24}(m+1) }{\eta^{24}(\tau)}   {\vf_{m}^{F}}(\tau) \, .
 \ee
 
 The nontrivial part of the decomposition theorem is that ${\vf_{m}^{F}}$ is a mock Jacobi form and hence admits a completion $ \widehat{\vf_{m}^{F}}$.
 The mock behavior can be summarized by the following partial differential equation
which is obeyed by its completion: 
\be\label{holanom}
 \tau_2^{3/2} \; \frac{\partial} {\partial \bar{\tau}}   \, \widehat{\vf_{m}^{F}}(\tau) \ = 
\sqrt{\frac{m}{8 \pi i}} \; \,  
 \sum_{\ell \, \mod \, (2m)}  {\overline{\vth_{m,\ell}(\tau)}} \, \vartheta_{m,\ell} (\tau,z) \, .
\ee
 
Thus,  puzzles concerning the degeneracies of supersymmetric quantum black holes  in string theory naturally lead to the mathematics of mock Jacobi forms. 

\section{Topology and Mock Modularity} 
\label{sec:topology}

%\subsection{Holomorphic Anomaly and Noncompactness}

As we explained earlier, the elliptic genus can be thought of as the right-moving Witten index, formally treating $\bar \tau$ as the right-moving inverse temperature. As a result, naively, this quantity is expected to be independent of $\bar \tau$ and is indeed so for a compact theory. The naive argument fails in a  non-compact conformal field theory, see e.g. \cite{Eguchi:2010cb,Troost:2010ud,Creutzig:2013hma,Murthy:2013mya,Harvey:2014nha,Dabholkar:2019nnc}. 
The holomorphic anomaly captures this `anomalous $\bar\tau$ dependence’ of the elliptic genus. To understand the physical origin of mock modularity, one would like to understand how precisely the holomorphic anomaly is related to the noncompactness of the underlying conformal field theory. 

To appreciate the essence of this phenomenon, one can consider a simpler problem about the  Witten index of a noncompact quantum mechanical system. In this context, the anomalous $\bar \t$ dependence that we are interested in can be  related to the anomalous $\b$ dependence of the Witten index where $\b$ is the inverse temperature. 

\subsection{Witten Index}
\label{sec:witten}

For a supersymmetric quantum field theory in a $d$-dimensional spacetime, the Witten index \cite{Witten:1982im} is defined by
\be\label{witten}
W(\b) := \Tr_\mathcal{H} \left[  (-1)^{F} e^{-\beta H} \right]
\ee 
where $H$ is the Hamiltonian, $F$ is the fermion number and $\cH$ is the Hilbert space of the theory. 
As usual, this trace can be related to a  supersymmetric  Euclidean path integral over a $d$-dimensional Euclidean base space $\Sigma$ with periodic boundary conditions for all fields schematically denoted by $\phi$  in the Euclidean time direction
%The Witten index  
% \eqref{witten} of  this worldline theory  has a path integral representation 
\begin{eqnarray}\label{witten2}
 W(\b) &=& \int d\phi \,  \langle  \phi |   (-1)^F e^{-\beta H} | \phi \rangle  
\qquad\qquad
 \nonumber\\
   &=& \int [d\phi] \, \exp\left(-S[\phi, \beta] \right) \, ,
\end{eqnarray}
where the path integral is over superfield configurations that are periodic in Euclidean time with period $\b$,  so the  Euclidean base space $\Sigma$ is a  circle of radius $\beta$ times the spatical manifold.

% \textbf{[PP: I think one should clarify something here, before $\Sigma$ was of arbitrary dimension $d$ and now it is a circle. Shouldn't it be a circle times a $(d-1)$-dimensional manifold?]}. 
%
%The Euclidean time $\t$ is related to the Lorentzian time $t$ as usual by Wick rotation $t = - i \tau$ and 
%the Euclidean action is:
%\begin{eqnarray}
%S[X, \b] = \frac{1}{2} \int_{0}^{\b} \, d\t  \left[  g_{ij }(x)\frac{dx^i }{d \t }\frac{dx^j }{d \t }
%+  \psi^a\left( \delta_{ab}
%\frac{d\psi^b }{d \t}
%+ \omega_{kab}
%\frac{dx^k }{d \t }
%\psi^b 
%\right)\right ]\, . 
%\end{eqnarray}
%The measure $[dX]$ is induced from the supermeasure\footnote{It is well-known that the supermeasure is  flat even if the  manifold $\cM$ is curved  because the factor of  $\sqrt{g}$ in the bosonic measure $dx := d^{2n}x \, \sqrt{g}$ cancels against a similar factor in the fermionic measure  
%$d\psi := d^{2n}\psi \,  \frac{1}{\sqrt{g}}$. }
% on the supermanifold $s\cM$. \textbf{[PP:  What are $\mathcal{M}$ and $X$?]}} 

If the quantum field theory is compact in the sense that the spectrum of the Hamiltonian is discrete, then  the  Witten index is independent of the inverse temperature $\b$:
\be\label{indWit}
\frac{d W(\b)}{d\b}  =0 \, .
\ee
This follows from the observation that the states with nonzero energy come in Bose-Fermi pairs and do not contribute to the Witten index \cite{Witten:1982im}. Only the zero energy states graded by $(-1)^F$ contribute and consequently, the Witten index is a  topological invariant. This is the case, for example, for a supersymmetric sigma model with a compact target space. 

By an appropriate choice of the sigma model, the Witten index in the zero temperature  ($\beta \rightarrow \infty$) limit can be related to some of the classic topological invariants such as the Euler character or the Dirac index of the target manifold.  Using temperature independence of the index, one can evaluate it in the much simpler high temperature   ($ \beta \rightarrow 0$) limit using the heat kernel expansion to prove the Atiyah-Singer index theorem \cite{Atiyah:1963zz}. Evaluating the path integral corresponding to the Witten index in this high-temperature semiclassical limit gives another derivation of the index theorem \cite{AlvarezGaume:1983rm,Friedan:1983xr}.

\subsection{Noncompact Witten Index}
\label{sec:noncompactwitten}

If the field space is noncompact and the spectrum is continuous, then the above argument can fail because now instead of a discrete indexed sum, one has an integral over a continuum of scattering states. To define the noncompact Witten index properly,  one needs a framework to incorporate the non-normalizable scattering states into the trace. One can address this issue  and give a suitable definition using the formalism of Gel’fand triplet \cite{gelfand}.

In general, the bosonic density of states in this continuum may not precisely cancel the fermionic density of states, and the noncompact Witten index can be temperature dependent. In other words, the equation \eqref{indWit} above can have an `anomaly’ in that the naive expectation expressed by \eqref{indWit}  can fail. It is possible to compute the temperature dependence using  localization of the supersymmetric path integral. The temperature dependent piece is no longer topological but is nevertheless  `semi-topological’ in that it is independent of any deformations that do not change the asymptotics. 

Some of the essential points about a noncompact path integral can be illustrated by a `worldpoint’  path integral where the   base space $\Sigma$ is a point and the  target space  $\mathcal{X}$ is a the real line $ -\infty <  u  <\infty $. We discuss this example first before proceeding to localization. 
The supersymmetric worldpoint action is given by 
\begin{eqnarray}
	S(u, F, \psi_-,\psi_+) =    \frac{1}{2} F^{2} + \iimg F \,  h'(u)
	+ \iimg h^{''}(u)\,  \psi_- \psi_+
\label{wpresult1}
\end{eqnarray}
where 
\be
h'(u) := \frac{dh}{du}\, , \qquad h^{''}(u) := \frac{d^{ 2}h}{du^{2}} \, . 
\ee
The path integral is now just an ordinary superintegral with flat measure\footnote{The  normalization factor $-\iimg$ can be understood as follows. Consider quantum mechanics of $2n$ real fermions. Denoting their zero-modes by $\psi_0^i,\;i=1,\ldots,2n$ we have:
\be
\bar\gamma = \iimg^{n} \gamma^{1} \dots\gamma^{2n} = (-2i )^{n} \psi^{1}_0\dots\psi_0^{2n}\qquad  \, .
\nonumber\ee
Moreover, $\Tr\,  \bar \gamma^{2} = 2^n$ which is the dimension of the spinor representation 
\be
\Tr\,  \bar \gamma^2 = N \int (-2i)^{n} \psi^{1}_0\dots\psi_0^{2n}  d\psi^{1}_0 \dots d\psi_0^{2n}
\nonumber
\ee
which implies $N= (-i)^{n}$. For two real fermions $n=1$ and  $N= -i$. }
\bea
W(\beta) = -\iimg \int_{-\infty}^{\infty} du \int_{-\infty}^{\infty} dF \int  d\psi_-\, d\psi_+ \, 
\exp\left[-\b S(U) \right] \, .
\eea

A particularly interesting  special case is
\be\label{htanh}
h'(u) = \l \tanh (a u) \, , 
\ee
for real $\l$. Integrating out the fermions and the auxiliary field $F$ gives
\bea
W(\beta) = -\sqrt{\frac{\beta}{2 \pi}}  \int_{-\infty}^\infty du\, 
h^{''}(u)\exp\left[- \frac{\beta}{2}(h'(u))^2 \right] 
\label{wpresult7}
\end{eqnarray}
 One can change variables 
\be
y = \sqrt{\frac{\b}{2}}h'(u)  \, , \quad  dy = \sqrt{\frac{\b}{2}}h^{''}(u) du
\ee
As $u$ goes from $-\infty$ to 
$\infty$, $y(u)$ is monotonically  increasing or decreasing  depending on if $\l$ is positive or negative; the inverse function $u(y)$ is single-valued,  and the integral  reduces to
\begin{eqnarray}
W(\beta) &=& -\frac{1}{\sqrt{\pi}} \int_{ -\sqrt{\frac{\b}{2}}\l}^{\sqrt{\frac{\b}{2}} \l} dy \, 
e^{-y^{2}} \, \nonumber \\
&=&  - \sgn (\l) \,  \erf \left(\sqrt{\frac{\b}{2}} |\l |\right) 
\label{wpresult8}
\end{eqnarray}
The  error function which appears naturally in this integral, has already made its appearance 
% in the proof of the APS theorem \cite{Atiyah:1975jf, Atiyah:1976jg} and also  
 in the  definition of the completion  \eqref{defstar} of a mock modular form in the guise of an incomplete Gamma function. 
Note that with a change of variable $y^{2} =t$, the integral 
\be
 \int_{ -\sqrt{\frac{\b}{2}}\l}^{\sqrt{\frac{\b}{2}} \l} dy \, 
e^{-y^{2}} \,
\ee
can be written as 
\be
\int_{0}^{\frac{\b}{2} \l^{2}} t^{-\half } \, e^{-t} \, dt
\ee
which we recognize as the lower incomplete gamma function 
$\gamma(\half, \frac{\b}{2} \lambda^{2})$ related to the upper incomplete gamma function \eqref{incomp}  by $\gamma(s, x) + \Gamma(s, x) = \Gamma(s)$. 

The noncompact Witten index is thus not temperature independent. The anomalous temperature dependence is captured by the equation
\be
\b^{\half} \frac{d}{d\beta} W(\b) \;\doteq\; \l \, \sgn(\l) \, e^{-\frac{\b \l^{2}}{2}}
\ee
which is a close analog of the holomorphic anomaly equation \eqref{ddtbarh}. Similar expressions appear in the path integral derivation of the holomorphic anomaly equation of mock modular forms encountered in gauge theory \cite{Dabholkar:2020fde}. 

The  worldpoint integral illustrates a number of important points. 
\begin{myenumerate}
\item
Without the fermionic integrations, the integral has a volume divergence because $h'(u)$ is bounded above for large $|u|$.  Inclusion of fermions effectively limits the integrand to the region close to the origin where $h'(u)$ varies,  and makes the integral  finite. 
%\item
%In the limit $\l \rightarrow 0$,  the action reduces to that of a free superparticle. In this case,  the integral is of the form $\infty \times 0$ and is ill-defined. Regularizing with $\l$ yields different answers depending on whether we approach $0$ from positive or negative side. This is related to the jump in the $\eta$-invariant when an eigenvalue  of the boundary operator $\cB$ crosses a zero in a spectral flow as explained before figure \textbf{\ref{fig5}}. 
\item
The answer depends only the asymptotic behavior of $h'(u)$ at $\pm \infty$ and is independent of any deformations that do not change the asymptotics. In  particular,  one would obtain the same result in  the limit $a \rightarrow \infty$ in \eqref{htanh}, when $h'(u)$ can be expressed in terms of the Heaviside step function:
\be 
h'(u) = \l \Big[\theta (u) - \theta (-u)\Big] \, .
\ee
\item
The appearance of the incomplete Gamma function in this context  is not a coincidence. The two turn out to be related through a path integral which localizes precisely onto the ordinary superintegral considered above.  For this reason, this example is particularly important for  understanding the connection between noncompactness and mock modularity. 
 
%
%Since $\bar \tau$ plays the role of right-moving temperature of the right-moving Witten index, one  can similarly relate this `anomalous temperature dependence’  arising from noncompactness to  mock modularity.
\end{myenumerate}

\section{Holomorphic Anomaly in  Gauge Theories}

In \cite{Vafa:1994tf} Vafa and Witten considered a topologically twisted version of maximally supersymmetric ($\mathcal{N}=4$) Yang-Mills theory in four dimensions. 
The partition function is naively expected to be a holomorphic function of the coupling constant $\tau$ and S-duality invariant. However,  the naive holomorphic partition function by itself is not S-duality invariant and  there is in fact a holomorphic anomaly as we describe below. 

\subsection{Vafa-Witten Partition Function}
\label{sec:4d-hol-anomaly}

The topological twist is defined by identifying  the $SU(2)= \Spin(3)$ subgroup of R-symmetry $SU(4)=\Spin(6)$ (with the standard embedding $\Spin(3)\subset \Spin(6)$) with the chiral $SU(2)$ subgroup of the $\Spin(4)\cong SU(2)\times SU(2)$ group of Euclidean rotations. The details of the twist are reviewed in the Section \ref{sec:6d-review} from the point of view of the parent 6d $\CN=(2,0)$ theory. We will denote by $Z^G[M^4,\tau]$ the partition function of the twisted theory with gauge group $G$ and complexified gauge coupling $\tau=\frac{4\pi i}{g^2}+\frac{\theta}{2\pi}$ on 4-manifold $M^4$. It has the following expected properties:
\begin{enumerate}[label=(\roman*)]
	\item $Z^G[M^4,\tau]\neq 0$ (generically, due to vanishing ghost number anomaly) 
	\item $Z^G[M^4,-1/\tau]=Z^{\hat{G}}[M^4,\tau]$ (up to a simple phase determined by 't Hooft anomalies) where $\hat{G}$ is the Langlands dual of $G$. E.g. $G=SU(2)$, $\hat{G}=PSU(2)= SO(3)$.
	\item (naively) $\frac{\partial}{\partial \bar\tau } Z^G[M^4,\tau] = 0$
	\item (naively) $\frac{\partial}{\partial g_{M^4} } Z^G[M^4,\tau] = 0$ where $g_{M^4}$ is the metric on $M^4$, that is the partition function is a topological invariant (only depends on the diffeomorphism class of the 4-manifold).
	\end{enumerate}  

The (naive) argument for both (iii) and (iv) is based on the fact that one can write the action of the twisted 4d SYM in the form
\begin{equation}
	S_\text{4d} =-i\frac{\tau}{8\pi^2}\int_{M^4} \mathrm{Tr} F \wedge F + Q(\Lambda)
\end{equation}
for some $\Lambda$, where $Q$ is one of the unbroken supercharges. The first term in the right-hand side is holomorphic in $\tau$ and independent of metric. Then, denoting by $\Phi$ the collection of all fields, we have in particular
\begin{equation}
	\frac{\partial}{\partial \bar\tau } Z^G[M^4,\tau]=
	\int D\Phi  \frac{\partial}{\partial \bar\tau } e^{-S_\text{4d}[\Phi]} = 
	\int D\Phi  Q\left(e^{-S_\text{4d}[\Phi]} \frac{\partial \Lambda}{\partial \bar\tau }\right) =0.
	\label{4d-stokes-naive}
\end{equation}
The last equality follows from the fact that $Q$ is a derivation on the space of fields, assuming there is no contribution from the boundary at infinity. Explicitly, if one introduces some coordinates $\Phi^i$ on the space of fields (e.g. harmonics on $M^4$) and $Q(\Phi^i)=\alpha^i[\Phi]$, then $Q=\sum_i \alpha^i[\Phi]\frac{\partial}{\partial \Phi^i}$ and one can apply Stokes theorem. However the assumption about vanishing contribution from infinity can in principle fail (and does in certain cases).

Moreover, under certain additional conditions (e.g. $M^4$ is K\"ahler with $c_1\geq 0$) the partition function can be localized on instanton solutions:
\begin{equation}
	Z^G[M^4,\tau] =\sum_{n} q^n\chi(\mathcal{M}_n^G[M^4])\qquad (q:=e^{2\pi i\tau}) 
	\label{VF-inst}
\end{equation}
where $\chi$ is the Euler characteristic and $\mathcal{M}_n^G[M^4]$ is the moduli space of $n$ $G$-instantons on $M^4$ (that is space of solutions to the equations $\ast F=-F$ with fixed $n=\frac{1}{8\pi^2}\int \mathrm{Tr} F\wedge F$, modulo gauge transformations). 

In one applies naively the formula (\ref{VF-inst}) for $M^4=\CP^2$ \cite{Klyachko:1991,Yoshioka:1994,Yoshioka:1995,Yoshioka:1996,Vafa:1994tf}:
\begin{eqnarray}
	Z^{SU(2)}_{\text{naive}}[\CP^2,\tau] & = & \frac{1}{2}f_0(\tau) \\
	 	Z^{SO(3)}_{\text{naive}}[\CP^2,\tau] & = & f_0(\tau)+f_1(\tau) 
\end{eqnarray}
where\footnote{Individual $f_v$ is the contribution from $SO(3)$ bundles with fixed $\Z_2$ flux $v=\int_{\mathbb{CP}^1} w_2$ through $\CP^1\subset \CP^2$. } $f_v(\tau)$, up to simple factors, are given by generating functions for Hurwitz class numbers $H(n)$ that have already appeared in Section \ref{sec:mock-example}: \begin{equation}
	f_v(\tau)=\sum_n\frac{3H(4n-v)q^{n-v/4}}{\eta(\tau)^3}\equiv \frac{3h_v(\tau)}{\eta(\tau)^3},\qquad v=0,1
\end{equation}
It turns out, that unlike in the case $M^4=K3$, the the property (ii) is not respected:
\begin{equation}
	Z^{SU(2)}_{\text{naive}}[\CP^2,-1/\tau] \neq 
	 	Z^{SO(3)}_{\text{naive}}[\CP^2,\tau]
\end{equation}
but ``almost''. Namely, $f_v(\tau)$ transform as a vector valued modular form that has a non-holomorphic modular completion:
\begin{equation}
	f_v(\tau)\rightsquigarrow f_v(\tau)+g_v(\tau,\bar\tau)
\end{equation}
where $g_v(\tau,\bar\tau)\rightarrow 0$ at $\bar\tau\rightarrow\infty$ and, as follows from (\ref{hol-anomaly-vector})-(\ref{hol-anomaly-vector-1}),
\begin{equation}
	\frac{\partial g_v(\tau,\bar\tau)}{\partial \bar\tau}=
	\frac{3}{\tau_2^{3/2}16\pi i \eta(\tau)^3}\sum_{n\in \Z} \bar{q}^{(n+v/2)^2}. 
	\label{g-hol-anomaly}
\end{equation}
Since the property (ii) is more fundamental than the property (iii), it is natural to guess that the modular completion above gives the full physical partition function. Its holomorphic anomaly is then determined by (\ref{g-hol-anomaly}). In particular
\begin{equation}
	\frac{\partial Z^{SU(2)}[M^4,\tau]}{\partial \bar\tau}=
	\frac{3}{\tau_2^{3/2}32\pi i \eta(\tau)^3}\sum_{n\in \Z} \bar{q}^{n^2}.
\end{equation}
By now the appearance of mock-modular forms in instanton counting problems has already a long history: \cite{1996alg.geom.12020G,Alexandrov:2016tnf,Alexandrov:2017qhn,Manschot:2017xcr,Korpas:2019cwg}.

\subsection{M-theory Realization}
\label{sec:6d-review}

In this section we briefly review some prerequisite facts about 6d maximally supersymmetric conformal field theories and their topologically twisted compactification on 4-manifolds. Six-dimensional $\mathcal{N}=(2,0)$ superconformal quantum field theories (QFTs) are known to be classified by Lie algebras of the form $\mathfrak{g}=\oplus_i \mathfrak{g}_i$, where each $\mathfrak{g}_i=\mathfrak{u}(1)$ or simply-laced (i.e. of ADE type). In these lectures by defult we assume Euclidean signature of space-time metric. The 6d $\CN=(2,0)$ extension of spin-Poincare symmetry (non-including conformal transformations) has $\Spin(6)_E \times_{\Z_2} \Spin(5)_R$ even subgroup, where $\Spin(6)_E$ is the symmetry of the spacetime rotations and $\Spin(5)_R$ is the R-symmetry. The extension has $16$ supecharges transforming in the representation $(\mathbf{4}_+,\mathbf{4})$.

Recall that $\Spin(n)$ group is the nontrival central extension of $SO(n)$ by $\Z_2$:
	\begin{equation}
		1\longrightarrow \Z_2 \longrightarrow \Spin(n) \longrightarrow SO(n) \longrightarrow 1.
	\end{equation}
Useful isomorphisms in low dimensions:
\begin{eqnarray}
	\Spin(3) & \cong & SU(2), \\
	\Spin(4) & \cong & \Spin(3)_\ell\times \Spin(3)_r \cong SU(2)_\ell \times SU(2)_r, \\
	\Spin(6) & \cong & SU(4).
\end{eqnarray}

Note that when $\mathfrak{g}=\mathfrak{u}(N)$ or $\mathfrak{su}(N)$ the  theory describes the worldwolume theory of $N$ M5-branes (respectively with or without center of mass degrees of freedom included) in 11-dimensional M-theory. The R-symmetry $\Spin(5)_R$ then can be understood as the group of rotations of the 5 normal directions.

The 6d $\CN=(0,2)$ theories do not have a Lagrangian description, however their compactification on a circle or a torus can be described in terms of maximally supersymmetric Yang-Mills gauge theory. In particular,	6d $\CN=(2,0)$ theory corresponding to a Lie algebra $\mathfrak{g}$, compactified on a 2-torus $T^2=\C/(\Z+\tau\Z)$ is equivalent to a 4d $\CN=4$ super-Yang-Mills (SYM) theory with gauge group $G$ such that $Lie(G)=\mathfrak{g}$ and the complexified coupling $\tau=\frac{4\pi i}{g^2}+\frac{\theta}{2\pi}$ equal to the complex structure of the torus.

Note that the choice of global structure of the gauge group $G$ is determined by the finite choice of ``sector'' of the 6d theory, which is a \textit{relative} QFT. It is similar to a 2d chiral Wess-Zumino-Witten conformal field theory (CFT), where one has to choose a conformal block. In the case of $\mathfrak{g}=\mathfrak{su}(N)$ there are two distinguished choices: $G=SU(N)$ and $G=PSU(N)=SU(N)/\Z_N$. They are related to each other by gauging a certain 1-form $\Z_N$ global symmetries.

Consider now a 6d $\CN=(2,0)$ theory on a spacetime of the form $M^4\times \Sigma^2$, where $M^4$ is a closed oriented 4-manifold and $\Sigma^2$ is a 2d spacetime. The group of local spacetime rotations then has a natural subgroup $\Spin(4)_E\times \Spin(2)_E \subset \Spin(6)_E$ where $\Spin(4)_E$ and $\Spin(2)_E$ are local rotations along $M^4$ and $\Sigma^2$ respectively. We will consider the theory topological twisted along $M^4$. In principle there are various ways to topologically twist the theory. The twist that we are interested in can be understood as identification of $\Spin(3)\subset \Spin(5)_R$ (embedded as the subgruop of rotations of 3 out 5 directions) with $SU(2)_\ell \subset \Spin(4)_E \cong SU(2)_\ell \times SU(2)_r$. Equivalently, this means turning on a background gauge field for $\Spin(3)$ subgroup of R-symmetry equal to $SU(2)_\ell$ component of the spin-connection on $M^4$.

When $\Sigma=T^2$, from the point of view of the 4d $\CN=4$ SYM on $M^4$ this twist is known as Vafa-Witten topological twist. The effective 4d theory is the Vafa-Witten theory \cite{Vafa:1994tf} so that its partition function $Z^G[M^4,\tau]$ considered in Section \ref{sec:4d-hol-anomaly} can be interpreted as the partition function of the 6d theory on $M^4\times T^2$.

\begin{exercise}
 \label{ex:reps-decomp}
	Consider the homomorphism
	\begin{equation}
		t:\;\Spin(4)'_E \times \Spin(2)_E
		\longrightarrow
		\Spin(6)_E \times \Spin(5)_R 
	\end{equation}
	correposning to the topological twist defined above. Namely, the component of the map in $\Spin(6)_E$ is given by the standard ebedding, while the component of the map in $\Spin(5)_R$ is given by the projection onto $SU(2)'_\ell \subset \Spin(4)'_E$ composed with the standard embedding $\Spin(3)\hookrightarrow \Spin(5)_R$. Show that the representations decompose as follows:
	\begin{eqnarray}
		t^*(\mathbf{4}_+,\mathbf{4}) & = &
		2(\mathbf{1},\mathbf{1})_{+\frac{1}{2}}
		+2(\mathbf{3},\mathbf{1})_{+\frac{1}{2}}
		+2(\mathbf{2},\mathbf{2})_{-\frac{1}{2}} \\
		t^*(\mathbf{1},\mathbf{5}) & = & 2(\mathbf{1},\mathbf{1})_0+ (\mathbf{3},\mathbf{1})_{0}
		\label{rep-decomp}
	\end{eqnarray}
	where in the right hand side the pair of numbers denote the dimensions of the representations of $SU(2)'_\ell \times SU(2)'_r$ and the index denotes the spin with respect to $\Spin(2)_E$.
\end{exercise}

Note that in the case when $\mathfrak{g}=\mathfrak{su}(N)$ and the 6d theory is the worldwolume theory of the stack of $N$ fivebranes in M-theory the twist can be realized geometrically in the following setup:
\begin{equation}	
			\begin{array}{cccccc}
				\text{M-theory} & \Lambda^{2+}TM^4 & \times  & \R^2 & \times & T^2 \\ 
				\text{$N$ M5's}  & M^4 &\times & \{0\} & \times & T^2  
			\end{array}
\end{equation}
where $\Lambda^{2+}TM^4$ is the rank 3 bundle of self-dual 2-forms over $M^4$.  This follows from (\ref{rep-decomp}) as the represenation $(\mathbf{3},\mathbf{1})$ can be realized by self-dual antisymmetric rank 2 tensors.

From the result of the Exercise \ref{ex:reps-decomp} it follows that the 6d theory twisted on $M^4$ has in general 2 unbroken supercharges. They both transform as right-moving spinors in the residual 2 dimensions. 

\subsection{The \texorpdfstring{$\mathfrak{su}(2)$}{su(2)} theory on the tensor branch}

For $\mathfrak{g}=\mathfrak{su}(2)$ the 6d theory is interacting and non-Lagrangian. However on the \textit{tensor branch} it can be effectively decribed in term of a single tensor multiplet corresponding to the Cartan subalgebra $\mathfrak{u}(1)\subset \mathfrak{su}(2)$ with extra interaction terms.

%\begin{remark}
	The tensor branch of the theory can be roughly understood as the phase where the scalar fields fluctuate over non-zero values. The non-compactness of the 6d theory can only arise in the directions corresponding to the large values of the scalar fields. Therefore to determine the holomorphic anomaly it should be enough to use the effective description of the theory on the tensor branch, which becomes asymptoticlly exact when the values of the scalar fields tend to infinity.
	In the M-theory setup the tensor branch corresponds to the phase when 2 fivebranes are separated in a normal direction.

The quantization conditions on the gauge transformations on the field $B$ are rescaled compared to the previous case\footnote{The normalization conventions are slighlty different from that of \cite{Dabholkar:2020fde}.}:
\begin{equation}
B\sim B+\beta,\; d\beta=0\;\;, \;\;\int_{\forall\text{ 2-cycle}}\beta\in 2\pi\sqrt{2} \Z.
\label{gauge-B-su2}
\end{equation}	
where the extra $\sqrt{2}$ factor is the length of the root in $\mathfrak{su}(2)$ algebra. Consequently, the fields of the effective 2d theory will be the same as in the $\mathfrak{g}=\mathfrak{u}(1)$ case, but with modified periodicity of the chiral scalars:
 \begin{equation}
	X_\pm^i\sim X_\pm^i + 2\sqrt{2}\pi n^i_\pm
\end{equation} 
where $n^i_\pm$ are quantized as before.	In the description in terms of a lattice CFT, this means that the lattice is now $\sqrt{2}H^2(M^4,\Z)$.

The 6d interaction term that will be relevant for us is \cite{Intriligator:2000eq,Ganor:1998ve}:
\begin{equation}
	\frac{i}{\sqrt{2}}\int B\wedge \eta_4
	\label{sky-term}
\end{equation}
Where $\eta_4$ is a 4-form that can be defined as follows. Let
\begin{equation}
	\hat\varphi^a:=\frac{\varphi^a}{\sqrt{\sum_{a}(\varphi^a)^2}},\qquad(D\hat\varphi)^a:=d\hat\varphi^a-A^{ab}_R\varphi^b, \qquad a,b=1,\ldots,5. 
\end{equation}
where $\varphi^a$ are the scalars of the tensor multiplet and $A^{ab}$ are the components of the background $\Spin(5)_R$ connection 1-form valued in the $\mathfrak{so}(5)$ algebra. Then
\begin{multline}
\eta_4:=\frac{1}{64\pi^2}\epsilon_{a_1a_2a_3a_4a_5}[
(D\hat{\varphi})^{a_1}(D\hat{\varphi})^{a_2}(D\hat{\varphi})^{a_3}(D\hat{\varphi})^{a_4}
\\
-2F^{a_1a_2}_R(D\hat{\varphi})^{a_3}(D\hat{\varphi})^{a_4}+F^{a_1a_2}_RF^{a_3a_4}_R]
\hat{\varphi}^{a_5}.
\label{eta4-form}
\end{multline}
where $F_R$ is the curvature of the R-symmetry connection $A_R$. The interaction term above is written in the untwisted theory, on a flat spacetime, but with non-trivial background $SO(5)$ R-symmetry gauge field.

Note that the term (\ref{sky-term}) corresponds to the fact that skyrmionic strings are electrically charged with respect to the $B$ field.	
In the dimensional reduction to 5d/4d gauge theory the corresponding term arises in 1-loop effective action on the Coulomb branch.

\section{Path integral derivation of holomorphic anomaly}

Our goal is to derive this holomorphic anomaly equation from the path integral. It is possible to do this from the path integral of the 4d SYM by calculating the non-vanishing boundary contribution in (\ref{4d-stokes-naive}) \cite{Dabholkar:2020fde}. However in these lectures we will use an alternative (but in many ways analogous) approach, also from \cite{Dabholkar:2020fde}.

\subsection{Reversing the order of compactification}

Instead of considering the 4d theory on $M^4$ obtained by compactification of the 6d theory on $T^2$ one can reverse the order of compactification and work with the 2d theory, usually denoted as $T_\mathfrak{g}[M^4]$ obtained by compactification of the 6d theory on $M^4$. The partition function of the Vafa-Witen theory on $M^4$ is equal to the partition function of this 2d theory on the torus $T^2$ with complex structure $\tau$ (with periodic-periodic boundary conditions on fermions). The latter can be also understood as a trace over the Hilbert space $\mathcal{H}$ of the 2d theory on a circle:
\begin{equation}
	Z^G[M^4,\tau]=Z_{T[M^4]}[T^2]=\mathrm{Tr}_{\mathcal{H}}(-1)^F q^{L_0}\bar{q}^{\bar{L}_0}.
\end{equation} 
As usual, $L_0=H+P$, $\bar{L}_0=H-P$ where $H$, $P$ are the Hamiltonian and momentum.  
The result of the Exercise \ref{ex:reps-decomp} tells us that the 2d theory $T_\mathfrak{g}[M^4]$ has at least $\mathcal{N}=(0,2)$ supersymmetry. Assuming the spectrum of the theory is discrete, the usual argument can be applied to show that the dependence on $\bar{q}$ in the trace above should drop out due to cancellation between femrionic and bosonic states with $\bar{L}_0\equiv Q^2\neq 0$ (where $Q$ is one of the right-moving supercharges). However, if the theory suffers from some sort of ``non-compactness'' and the spectrum is continuous, the canecellation can fail and the partition function can have an anomalous dependence on $\bar{q}$. As we will see, this is indeed the case for $T_{\mathfrak{su}(2)}[\CP^2]$.

%%%%%%%%%%%%%%%%%%%

To summarize, the computations of the  holomorphic anomaly both in the 2d sigma model description and directly in the 4d gauge theory description give the same result  on the right hand side of  (\ref{g-hol-anomaly}), but the various factors have different origins in the two regions.  
\begin{myitemize}
\item The  factor of $3$  is related to the to the  quantum of $H$-flux in the sigma-model description and to the first Chern class of the canonical line bundle of $\mathbb{CP}^{2}$ in the gauge theory description.
\item  The factor of  $\tau_2^{-3/2}$ comes   from the integral over three non-compact bosonic zero-modes in the sigma model description and from the integral over the constant mode of the auxiliary field in the gauge theory description.
\item  The factor of  $\eta(\tau)^{-3}=\eta(\tau)^{-\chi(\mathbb{CP}^2)}$ is the contribution of the left-moving oscillators in the sigma model description and of point-like instantons in the gauge theory description.
 \item Finally, the anti-holomorphic theta-function $\sum_{n\in\mathbb{Z}}\bar{q}^{n^2}$ is a contribution of  right-moving momenta of a  compact chiral boson in the sigma model description and of abelian anti-instantons in the gauge theory decsription. 
\end{myitemize}

We now comment on the relation between the holomorphic anomaly and mock modularity. The naive holomorphic partition function of the twisted $SO(3)$ super Yang-Mills theory on $\mathbb{CP}^{2}$  is the  holomorphic kernel of the  anomaly equation (\ref{g-hol-anomaly}) which receives contributions only from the instantons. It is holomorphic but not modular.   The presence of the  holomorphic anomaly implies that  the physical partition function  necessarily contains a nonholomorphic piece given by an Eichler integral of the anomaly \cite{Zagier:1975a} which receives contributions from the anti-instantons.  In the modern terminology \cite{Zwegers:2008zna,MR2605321,Dabholkar:2012nd} discussed earlier, 
the holomorphic piece is  a (vector valued)  `mixed mock modular form’ whereas the   anomaly is governed by its `shadow’. The physical partition function satisfying the anomaly equation is the `modular completion’ and has  good modular properties, as expected from duality.

These considerations extend naturally  to other K\"ahler 4-manifolds with $b_{2}^{+}=1$, $b_1=0$ and to other groups \cite{Dabholkar:2020fde}. In general, when  the configuration space of the twisted theory is noncompact, the partition function is modular but not holomorphic, and satisfies a holomorphic anomaly equation similar to (\ref{g-hol-anomaly}).  As we have seen, this incompatibility  between holomorphy and modularity is the essence of mock modularity.  The  physical requirement of duality invariance of the  path integral  thus leads naturally to the mathematical formalism of mock modularity whenever the relevant configuration space is noncompact.

%%%%%%%%%%%%%%%%%

Before we proceed with the derivation of the holomorphic anomaly equation from the point of view of the effective 2d theory for $\mathfrak{g}=\mathfrak{su}(2)$, consider first the case of $\mathfrak{g}=\mathfrak{u}(1)$. The corresponding 6d theory is the free theory of $\mathcal{N}=(2,0)$ tensor multiplet. It consists of the fields forming the following rerpresentations of $\Spin(6)_E \times \Spin(5)_R$:
\begin{enumerate}[label=(\alph*)]
	\item scalars $\varphi$ in $(\mathbf{1},\mathbf{5})$
	\item fermions in  $(\mathbf{4}_+,\mathbf{4})$
	\item a self-dual 2-form field $B$  in $(\mathbf{15},\mathbf{1})$ ($dB=\ast dB$). In general it is only locally well-defined and is subject to 1-form gauge transformations:
\begin{equation}
B\sim B+\beta,\; d\beta=0\;\; (\beta=d\alpha\text{ locally}), \;\;\int_{\forall\text{ 2-cycle}}\beta\in 2\pi \Z.
\label{gauge-B}
\end{equation}

\end{enumerate}

In what follows we will assume that $M^4$ is simply-connected (in particular $b_1=0$). Then $H^2(M^4,\Z)$ can be identified with subgroup of de Rham cohomology $H^2(M^4,\R)$ represented by 2-forms $h$ such that $\int_{\text{a 2-cycle}}h\in \Z$. As usual, we will denote by $b_2^\pm$ the number of positive/negative eigenvalues of the bilinear form on $H^2(M^4,\Z)$ given by $\int_{M^4} \cdot \wedge \cdot$. 

  The effictive 3d theory $T_{\mathfrak{u}(1)}[M^4]$ can be obtained by the standard Kaluza-Klein (KK) reduction, by counting harmonic forms on $M^4$ and taking into account the results of the Exercise \ref{ex:reps-decomp} (cf. \cite{Dedushenko:2017tdw,Gadde:2013sca}):
\begin{enumerate}[label=(\alph*)]
	\item $2+b_2^+$ real scalars $\phi^i$
	\item $2+2b_2^+$ Majorana-Weyl right-moving fermions 
	\item $b_2^\pm$ right/left-moving chiral compact scalars $X_\pm^i$.
\end{enumerate}
The result (c) can be seen as follows. The (massless) KK reduction of the field $B$ is given by the following decomposition:
\begin{equation}
	B=\sum_{i=1}^{b_2^+} X^i_+\,h_i^+ + \sum_{i=1}^{b_2^-} X^i_-\,h_i^-
\end{equation}
where $h^\pm_i$ are the basis in the space of harmonic 2-forms on $M^4$ chosen such that
\begin{eqnarray}
	\ast_\text{4d} h_i^\pm & = & \pm h_i^\pm, \\
	\int_{M^4} h_i^+h_j^- & = & 0, \\
	\int_{M^4} h_i^\pm h_j^\pm & = & \pm\delta_{ij} .
\end{eqnarray}
From the self-duality $dB=\ast_\text{6d} dB$ it then follows that $\ast_\text{2d}dX^i_\pm = \pm dX^i_\pm$. Equivalently, using a complex coordinate $z$ to parametrize the 2d spacetime, $\partial X_+^i=0$, $\bar\partial X_-^i=0$. From the large gauge transformations (\ref{gauge-B}) it follows that $X_\pm^i$ are compact:
\begin{equation}
	X_\pm^i\sim X_\pm^i + 2\pi n^i_\pm
\end{equation} 
where $n^i_\pm$ are quantized such that $\sum_{i,\pm} n^i_\pm h_i^\pm \in H^2(M^4,\Z)$. That is $X_\pm^i$ can be understood as components of a field valued in the $b_2$-torus $H^2(M^4,\mathbb{R})/2\pi H^2(M^4,\Z)$.

	The collection of field $X^i_\pm$ form what is usually called Narain's lattice CFT corresponding to the indefinite lattice $H^2(M^4,\Z)\cong H_2(M^4,\Z)$ (for closed $M^4$, as we assume, the lattice is self-dual) with the bilinear form $\int_{M^4} \cdot \wedge \cdot$. 

\begin{exercise}
\label{ex:kaehler-susy}
	Show that when $M^4$ is (hyper-)K\"ahler, that is the holonomy is reduced to $(SU(2))\;U(2)\subset SO(4)_E$, there are (8)\,4 unbroken supercharges, all right-moving in 2d.
\end{exercise} 

\subsection{Review of 2d \texorpdfstring{$\mathcal{N}=(0,1)$}{minimally supersymmetric} sigma-models}

For our purposes it is enough to consider $\mathcal{N}=(0,1)$ sigma-models without left-moving fermions. The theories from such class are defined by the choice of a \textit{target} manifold $\mathcal{X}$ equipped with Riemannian metric and a closed 3-form $h\in \Omega^3(\mathcal{X})$. The target manifold should satisfy certain topological constraints for theory to be well defined on the quantum level ($w_1(T\mathcal{X})=0$, $w_2(T\mathcal{X})=0$, $\frac{1}{2}p_1(T\mathcal{X})=0$). Let $\Sigma$ be the 2-dimensional spacetime (also known as \textit{source} or \textit{worldsheet}). The field content of the theory is then the following:
\begin{itemize}
	\item scalars $\phi:\Sigma \rightarrow \mathcal{X}$,
	\item fermions $\psi \in \Gamma(\phi^*(T\mathcal{X})\otimes S_+\Sigma)$
\end{itemize}
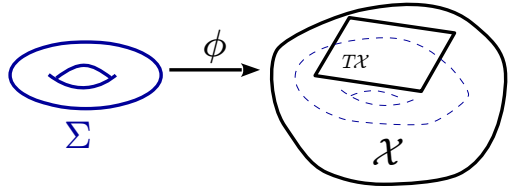
\begin{center}
	\includegraphics[scale=1.8]{01-sigma-model-map-simple}
\end{center}
where $S_+$ it the bundle of right-moving chiral spinors on $\Sigma$. Informally, the fermion fields can be understood as right-moving spinors valued in the tangent space of the target (pulled back to the 2d spacetime). In \textit{local} coordinates on the target, and assuming that $\Sigma$ is flat, the action reads \cite{Hull:1985jv,Brooks:1986uh,melnikov2019introduction}:
\begin{eqnarray}
    S_\text{2d} & = \cfrac{1}{4\pi} \int d^2z\left((g_{ij}(\phi)+b_{ij}(\phi))\partial \phi^i\bar \partial\phi^j +g_{ij}\psi^i\partial\psi^j-(\Gamma_{ijk}+\frac{1}{2}h_{ijk})\psi^k\psi^i\partial\phi^j\right)
    \label{NLSM-action}
\end{eqnarray}
where $d^2z:=idzd\bar{z}$, $\Gamma_{ijk}$ are Christoffel symbols of the Levi-Cevita connection, and $b_{ij}$ are the components of a 2-form $b\in \Omega^2(\mathcal{X})$ such that $h=db$. The term containing the scalar fields and the 2-form $b$  in the target is the Wess-Zumino term, which can be recast as 
\begin{equation}
    S_\text{2d WZ}=\frac{i}{4\pi}\int_{\Sigma^2} \phi^*(b) =\frac{i}{4\pi}\int_{\Xi^3} \phi^*(h)
	\label{2d-WZ-term}
\end{equation}
where $\partial \Xi^3=\Sigma^2$. The supercharge acts on the fields as follows:
\begin{equation}
    \begin{array}{rcl}
        [Q,\phi^i] & = & \psi^i,  \\
         \{Q,\psi^i\}& = & -\bar\partial \phi^i .
    \end{array}
    \label{Q-2d}
\end{equation}
The right-moving energy momentum-tensor and the supercurrent (i.e. the Noether current for the symmetry $Q$) are given by (cf. \cite{delaOssa:2018azc}) 
\begin{eqnarray}
        \bar{T} & = & 
        -\frac{1}{2}\,g_{ij}\bar\partial \phi^i\bar\partial \phi^j
        -\frac{1}{2}\,g_{ij}\psi^i\bar\partial \psi^j
        +\frac{1}{2}\,\partial_kg_{ij}\psi^k\psi^j\bar\partial\phi^i
        +\frac{1}{4}\,h_{ijk}\psi^i\psi^j\bar\partial\phi^k
        ,  \\
         \bar G& = & i\left(g_{ij}\psi^i\bar\partial\phi^j-\frac{1}{3!}\,h_{ijk}\psi^i\psi^j\psi^k\right). 
		 \label{sigma-super-current}
 \end{eqnarray}
By definition $\bar{T}\equiv T_{\bar{z}\bar{z}}=2\pi \delta S_\text{2d}/\delta h_{\bar z \bar z}$, where $h_{\bar z \bar z}$ is the corresponding component of the metric on the 2d space-time. 

\begin{exercise}
\label{ex:supercurrent-check}
Check that the supercurrent $\bar{G}$ above satisfies 
\begin{equation}
    \{Q,\bar{G} \} = 2i\,\bar{T}.
    \label{G-T-relation}
\end{equation}
\end{exercise}

The path integral for partition function of the 2d theory (which we will denote as $\sigma(\mathcal{X})$) on a 2-torus with periodic-periodic boundary conditions on fermions localizes on constant maps to the target. If the target $\mathcal{X}$ is a closed manifold ($\partial \mathcal{X}=0$) the result has an explicit formula\footnote{As usual, it is assumed that $\int_\mathcal{X} f\equiv 0$ for $f\in\Omega^n(\mathcal{X}),\, n\neq \mathrm{dim}\mathcal{X}$.} \cite{Witten:1986bf}:
\begin{equation}
	Z_{\sigma(\mathcal{X})}[T^2] = \mathrm{Tr}_\mathcal{H} (-1)^F q^{L_0}\bar{q}^{\bar{L}_0} = \int_\mathcal{X} \sqrt{\det\frac{\mathcal{R}/2\pi i}{\theta(\mathcal{R}/2\pi i;\tau)}}
	\label{EG-compact}
\end{equation}
where $\mathcal{R}\in \Omega^2 \otimes \mathfrak{so}(\mathrm{dim}\mathcal{X})$ is the curvature 2-form on $\mathcal{X}$ and 
\begin{equation}
	\theta(u;\tau):=q^{\frac{1}{12}}(e^{u/2}-e^{-u/2})\prod_{n\geq 1} (1-q^ne^u)(1-q^ne^{-u})\equiv \frac{i\theta_1(\tau,\frac{u}{2\pi i})}{\eta(\tau)}.
\end{equation}

The characteristic class of $\mathcal{X}$ in the right-hand side of (\ref{EG-compact}) is known as \textit{elliptic genus}.

\begin{exercise}
\label{ex:elliptic-genus-char-class}
	Show that
	\begin{equation}
	\int_\mathcal{X} \sqrt{\det\frac{\mathcal{R}/2\pi i}{\theta(\mathcal{R}/2\pi i;\tau)}} =
	\frac{1}{\eta(\tau)^{\mathrm{dim}\mathcal{X}}} \int_\mathcal{X}
	\left\{
	1+E_2(\tau)p_1+\left(\frac{E_2(\tau)^2}{2}p_1^2+\frac{E_4(\tau)}{12}(p_1^2-2p_2)\right)+\ldots
	\right\},
	\end{equation}
	where $p_i$ are Pontryagin classes of $\mathcal{X}$ and find the next term. The $n$-th term in the expansion is a characteristic class of degree $4n$ with coefficients in being quasi-modular forms of $SL(2,\Z)$ (i.e. polynomials in the Eisenstein series $E_{2k}$) of weight $2n$. Argue that when $p_1=0$ the dependence on $E_2$ drops out.
\end{exercise}

If $\mathcal{X}$ is non-compact, however, $\frac{\partial Z_{\sigma(\mathcal{X})}}{\partial \bar\tau} \neq 0$ in general. Assuming that $\mathcal{X}$ asymptotically is of the form $\mathcal{Y}\times \R$ (informally $\partial \mathcal{X} = \mathcal{Y}$), it was argued in \cite{Gaiotto:2019gef} that the anti-holomorphic derivative can be expressed in terms of a 1-point function of the supercurrent of in the sigma model with target $\mathcal{Y}$. Namely
	\begin{equation}
		\frac{\partial Z_{\sigma(\mathcal{X})}[T^2]}{\partial \bar\tau}=
		\frac{-e^{\pi i/4}}{\sqrt{8\tau_2}\eta(\tau)}\langle \bar{G}\rangle_{\sigma(\mathcal{Y})}
	\end{equation} 
where $\langle \ldots \rangle_{\mathcal{T}}$ denotes a 1-point function in a theory $\mathcal{T}$ on $T^2$.

Below we present a sketch of the proof of this formula and address to \cite{Gaiotto:2019gef,Dabholkar:2020fde} for details. Omitting numerical factors we have
\begin{equation}
	\frac{\partial Z_{\sigma(\mathcal{X})}[T^2]}{\partial \bar\tau} \propto
	\int D\phi D\psi \frac{\partial}{\partial h_{\bar{z}\bar{z}}} e^{-S_\text{2d}[\phi,\psi]} \propto
	\langle \bar{T}\rangle_{\sigma(\mathcal{X})}
	\propto
	\langle \{ Q,\bar{G}\}\rangle_{\sigma(\mathcal{X})}
	\propto
	\frac{1}{\sqrt{\tau_2}\eta(\tau)}\langle \bar{G}\rangle_{\sigma(\mathcal{Y})}
	\label{boundary-sketch}
\end{equation}
The last equality is the most non-trivial and can be argued as follows. A correlator of a $Q$-invariant observable on the torus localizes on the space of zero-modes, that is space of constant functions $\phi$ and $\psi$ on $T^2$:
\begin{equation}
	\langle \{ Q,\bar{G}\}\rangle_{\sigma(\mathcal{X})} = \int_{\text{zero modes}} d\phi_0 d\psi_0 f(\phi_0,\psi_0)
\end{equation}
where $f(\phi_0,\psi_0)$ is result of the integration over nonzero bosonic and fermions modes. The functions on the space of bosonic and femrmionic zero-modes are in 1-to-1 correspondence with forms on the target $\mathcal{X}$. Namely, the components of the bosonic zero-modes can be identified with local coordinates $\phi_0^i$ on $\mathcal{X}$ and the components of the fermionic zero-modes can be identified with their differentials: $\psi_0^i \equiv d\phi_0^i$. Under this correpondence
\begin{equation}
	\int_\text{zero modes} d\phi_0 d\psi_0 f(\phi_0,\psi_0) = \int_\mathcal{X} f
\end{equation}
where in the right-hand side $f$ is understood as the corresponding element of $\Omega^\ast(\mathcal{X})$.  Moreover, the supercharge (\ref{Q-2d}) restricted to the space of zero modes can be indentified with the exterior derivative under this correspondence:
\begin{equation}
	Q|_\text{zm}=\sum_i \psi^i_0\frac{\partial }{\partial \phi_0^i} \equiv d\qquad (\text{acting on }\Omega^*(\mathcal{X}))
\end{equation}
Therefore $f=dg$ where $g$ corresponds to $\bar{G}$ in the way $f$ corresponds to $\bar{T}$ and 
\begin{equation}
	\int_\mathcal{X} f = \int_\mathcal{Y} g
\end{equation} 
from Stokes theorem. The extra factor in (\ref{boundary-sketch}) is the contribution of bosonic and fermionic nonzero modes corresponding to the extra direction $\R$ in $\mathcal{X}$:
\begin{equation}
	\frac{1}{\sqrt{\tau_2}|\eta(\tau)|^2}\cdot \overline{\eta(\tau)} = \frac{1}{\sqrt{\tau_2}\eta(\tau)}
\end{equation}

\subsection{Holomorphic anomaly for \texorpdfstring{$\CP^2$}{complex projective space}}

Consider now the particular case of $M^4=\CP^2$. Although the final result is independent on the choice of metric, assume the standard Fubini-Studi metric (which is K\"ahler) for concreteness. The group $H^2(\CP^2,\Z)$ has a single generator represented by the self-dual form $h_+^1=\omega$ equal to the K\"ahler form, satisfying $\int_{\CP^2} \omega^2 =1$ (also $\int_{\CP^1} \omega=1$). In particular $b_2^+=1$ and $b_2^-=0$. According to the general result, the field content of the effective 2d theory $T_{\mathfrak{su(2)}}[\CP^2]$ on the tensor branch is then the following:
\begin{enumerate}[label=(\alph*)]
	\item 3 scalars $\phi^i,\;i=1,2,3$, 
	\item 4 fermions $\psi^i,\;i=1,2,3,4$,
	\item 1 right-moving compact chiral boson $X_+$ of radius $\sqrt{2}$. 
\end{enumerate}
This theory is a sigma-model with target $\mathcal{X} = \R^3\times S^1$ with the exception that the scalar field valued in $S^1$ is chiral right-moving. The missing left-moving component, however is not participating in supersymmetry tranformations anyway. Topologically this target space looks like $S^2\times S^1 \times \mathbb{R}$ near infinity, where $\R$ is the radial direction of $\mathbb{R}^3$.

\begin{exercise}
\label{ex:NS-flux}
	Show that the 6d skyrmionic string term (\ref{sky-term}) reduces to a 2d Wess-Zumino term (\ref{2d-WZ-term}) with 
	\begin{equation}
		h=3\cdot 8\pi^2 \cdot \text{vol}_{S^2} \wedge \text{vol}_{S^1}
	\end{equation}
	where $\text{vol}_{S^2}$ and $\text{vol}_{S^1}$ are the standard volume forms on $S^2$ and $S^1$ normalized so that the total volume is $1$. Use the fact that one can always choose $SO(5)_R$ indices so that the 2d fields $\phi^i,\,i=1,2,3$ arise as constant modes of $\varphi^a,\,a=1,2,3$. Then in the twisted theory $F_R^{ab}=0$ unless $(a,b)$ is a permutation of $(4,5)$, while $F^{45}_R/(2\pi)$ is equal to (up to a sign) to the Ricci form, which is $3\omega$ for $\CP^2$. 
\end{exercise}

To calculate the holomorphic anomaly it is then enough to calculat the 1-point function of the supercurrent given by (\ref{sigma-super-current}) in the sigma-model with target $\mathcal{Y}=S^2\times S^1$:
\begin{equation}
	\langle \bar{G} \rangle_{\sigma(\mathcal{X})} = 
	\langle ig_{ij}\psi^i\bar\partial\phi^j-\frac{i}{3!}\,h_{ijk}\psi^i\psi^j\psi^k \rangle_{\sigma(\mathcal{X})}
\end{equation}
One can argue that the first term in the correlator does not contribute, as it cannot saturate the integral over 3 fermionic zero-modes. The correlator of the second term, as before, localizes to the integral over zero-modes with an overall factor from the contribution of non-zero modes. Ommitting non-essential numerical factors (for a careful calculation see \cite{Dabholkar:2020fde}), the integral over zero-modes reads
\begin{equation}
	\int_{S^2\times S^1} h = 3\cdot 8\pi^2
\end{equation}
where we used again the correspondence between functions of zero-modes and forms on the target. The contribution from nonzero modes is
\begin{equation}
\overline{\eta(\tau)^3} 
\cdot 
\frac{1}{\tau_2 \eta(\tau)^2\overline{\eta(\tau)^2}}\cdot \frac{\sum_{n}\bar{q}^{n^2}}{\overline{\eta(\tau)}}
\end{equation}
where the first factor is the contribution of nonzero modes of 3 right-moving fermions, the second factor is the contribution of nonzero modes of 2 scalars valued in $S^2$, and the final factor is the contribution of the chiral right-moving bosons of radius $\sqrt{2}$ (in the vacuum sector). Combining the contributions from zero- and nonzero modes together we arive at the desired formula for the holomorphic anomaly:
\begin{equation}
	\frac{\partial Z^{SU(2)}[M^4,\tau]}{\partial \bar\tau}=
	\frac{3}{\tau_2^{3/2}32\pi i \eta(\tau)^3}\sum_{n\in \Z} \bar{q}^{n^2}.
\end{equation}

	The choice of the vacuum sector in the theory of chiral compact boson of radius $\sqrt{2}$ correponds to the choice of the gauge group $G=SU(2)$ (equivalently, up to a factor, the sector of $SO(3)$ gauge theory with $\int_{\CP^1} w_2=0$). The other sector (out of 2 in total) of the chiral boson corresponds to the sector of $SO(3)$ gauge theory with $\int_{\CP^1} w_2=1$ and gives the result for the right-hand side of (\ref{g-hol-anomaly}) for $v=1$.

\appendix

\section{Solutions}

\paragraph{Exercise \ref{ex:cusp}.}

A weight $12$ modular form must be a linear combination  of the form \eqref{weight12}. For a  cusp form, the $q$-expansion has no constant term. Hence we determine $a=-b$ to  prove that 
\be
\Delta(\tau) = \frac{1}{1728} \left( E_{4}^{3} (\t) - E_{6}^{2} (\t) \right)
\ee
is the unique weight $12$ cusp form, normalized so that  the $q$-expansion starts with $q$.

\paragraph{Exercise \ref{ex:T4genus}.}

The SCFT with target space $T^{4}$ has four free bosons and four free fermions; hence the total central charge is $6$. The path integral for the elliptic genus vanishes because  the right-moving fermions with the periodic boundary condition in the Ramond sector have zero modes. To get a nonzero answer one can define a modified elliptic genus $\chi_{mod}(\tau, z|\mathcal{X})$ with two insertions of the current  $J_{R}$ in the trace \eqref{elliptic-def} to soak up the four fermion zero modes. Since the current operator has conformal weight $1$, now the path integral  transforms with weight $-2$ rather than with weight $0$. This implies that the elliptic genus should be a Jacobi form of index $1$ and weight $-2$ and hence must be proportional to $A(\t, z)$. 
It is easy to confirm this by an explicit computation for the free bosons and fermions using the product representation of the theta function which also fixes the overall normalization.

\paragraph{Exercise \ref{ex:K3genus}.}

The SCFT with target space $K3$ also has central charge $6$. This is easy to see in the large volume limit where the curvarture can be ignored and one has four free bosons and four free fermions.  This implies that the elliptic genus of $K3$ must be a Jacobi form of index $1$ and weight $0$; and hence must be proportional to $B(\tau, z)$. To fix the normalization, we use the fact that when $z=0$ the elliptic genus reduces to the Witten index which gives the Euler number of $K3$ which is $24$. Hence, the elliptic genus of $K3$ equals $B(\tau, z)$. 

This general conclusion can be verified by an explicit computation at a point  in the moduli space where the SCFT is exactly solvable. For example, one can consider the point where the $K3$ is represented as an orbifold of a 4-torus $T^{4}/\mathbb{Z}_{2}$ where the $\mathbb{Z}_{2}$ symmetry with generator $R$ reflects the bosonic coordinates of the torus and their fermionic partners (see, for example, \cite{Dabholkar:1997zd}). The elliptic genus \eqref{elliptic-def} can then be written as a projected trace over untwisted (U) and twisted (T) sectors as 
\be
\label{elliptic-orb}
\chi (\tau, z| K3) =  \half \mathrm{Tr}_{U} (1 + R) + \half \mathrm{Tr}_{T} (1 + R) \,  \, ,
\ee
(with all other operators implicitly in the trace as in \eqref{elliptic-def}). 
The first term vanishes because of the fermionic zero modes as explained above. The second term gives the first term in \eqref{B}. There are 16 fixed points of the orbifolding symmetry and in the twisted sector the  fields are half-integer moded.  Summing over the contributions from all fixed points, the  third term gives the second term in \eqref{B} whereas the fourth term gives the third term in \eqref{B} as can be seen from the product representation of the theta functions. 

\paragraph{Exercise \ref{ex:reps-decomp}.}

First we decompose the chiral spinor representation of $\Spin(6)_{E}$ into a sum of tersor products of chiral spinor represntations of $\Spin(4)_E=SU(2)_\ell\times SU(2)_r$ and $\Spin(2)_E$ with respect to the standard emebedding $\Spin(4)_E\times \Spin(2)_E\subset \Spin(6)_E$ (with two chiralities correlated):
\begin{equation}
    \mathbf{4}_+=(\mathbf{2},\mathbf{1})_{+\frac{1}{2}}+(\mathbf{1},\mathbf{2})_{-\frac{1}{2}}.
\end{equation}
The non-chiral spinor representation of $\Spin(5)_R$ splits into two copies of non-chiral spinor representation of $\Spin(3)_R\subset \Spin(5)_R$:
\begin{equation}
    \mathbf{4}=2\cdot\mathbf{2}.
\end{equation}
Taking their product and identifying $SU(2)_\ell$ with $\Spin(3)_R$ we have:
\begin{equation}
    \mathbf{4}_+ \otimes \mathbf{4} =
    2\cdot(\mathbf{2}\otimes \mathbf{2},\mathbf{1})_{+\frac{1}{2}}+2\cdot (\mathbf{2}\otimes \mathbf{1},\mathbf{2})_{-\frac{1}{2}}
    =
    2\cdot(\mathbf{3},\mathbf{1})_{+\frac{1}{2}}+2\cdot(\mathbf{1},\mathbf{1})_{+\frac{1}{2}}+2\cdot (\mathbf{2},\mathbf{2})_{-\frac{1}{2}}.
\end{equation}
Similarly, the vector representation of $\Spin(5)_R$ splits into the vector two scalar representations of $\Spin(3)_R$:
\begin{equation}
    \mathbf{5}=\mathbf{3}+2\cdot\mathbf{1}
\end{equation}
so that
\begin{equation}
    \mathbf{1}\otimes \mathbf{5}=
     (\mathbf{3},\mathbf{1})_{0}+2\cdot ( \mathbf{1},\mathbf{1})_{0}.
\end{equation}

\paragraph{Exercise \ref{ex:kaehler-susy}.}
In the K\"ahler case the holonomy is reduced down to the subgroup $U(1)_\ell' \times SU(2)_r'\subset \Spin(4)_E'$. The right-hand side of the first equation in (\ref{rep-decomp}) then decomposes as follows:
\begin{equation}
     4\cdot \mathbf{1}_{0,+\frac{1}{2}}+2\cdot \mathbf{1}_{+1,+\frac{1}{2}}+2\cdot \mathbf{1}_{-1,+\frac{1}{2}}+2\cdot \mathbf{2}_{+\frac{1}{2},-\frac{1}{2}}+2\cdot \mathbf{2}_{-\frac{1}{2},-\frac{1}{2}}.
\end{equation}
where the bold numbers denote irreducible representations of $SU(2)_r'$ of the corresponding dimensions and the subscripts denote the charges of $U(1)_\ell'\times \Spin(2)_E$. This indeed contains four scalars on $M^4$ (the first term in the decomposition), all of which are right-moving on $\Sigma$.

On a hyper-K\"ahler $M^4$ the group $U(1)'_\ell$ is reduced further to the trivial one, so that we have the following decomposition into representations of $SU(2)_r'\times \Spin(2)_E$:
\begin{equation}
     8\cdot \mathbf{1}_{+\frac{1}{2}}+4\cdot \mathbf{2}_{-\frac{1}{2}}.
\end{equation}
This indeed has 8 scalars on $M^4$, all of which are right-moving on $\Sigma$.

\paragraph{Exercise \ref{ex:supercurrent-check}.}
Acting by $Q$ on the first term of $\bar{G}$ and using (\ref{Q-2d}) we have
\begin{multline}
    \{Q,ig_{ij}\psi^i\bar\partial\phi^j\}=
    i[Q,g_{ij}]\psi^i\bar\partial\phi^j+
    ig_{ij}\{Q,\psi^i\}\bar\partial\phi^j-
    ig_{ij}\psi \bar\partial[Q,\phi^j]=\\
        i\partial_k g_{ij}\psi^k\psi^i\bar\partial\phi^j
    -ig_{ij}\bar\partial \phi^i\bar\partial\phi^j-
    ig_{ij}\psi \bar\partial \psi^j.
\end{multline}
Similarly, action on the second term gives
\begin{multline}
    \{Q,-\frac{i}{3!}h_{ijk}\psi^i\psi^j\psi^k\}=
    -\frac{i}{3!}[Q,h_{ijk}]\psi^i\psi^j\psi^k
    -\frac{i}{2}h_{ijk}\psi^i\psi^j\{Q,\psi^k\}=\\
    -\frac{i}{3!}\frac{\partial{h_{ijk}}}{\partial\phi^r}
    \psi^r\psi^i\psi^j\psi^k+
    \frac{i}{2}h_{ijk}\psi^i\psi^j\bar\partial\phi^k=
    0+\frac{i}{2}h_{ijk}\psi^i\psi^j\bar\partial\phi^k
\end{multline}
where we used closedness of the 3-form $h$. Combining these together we indeed arrive at (\ref{G-T-relation}).

\paragraph{Exercise \ref{ex:elliptic-genus-char-class}.}

Recall the following identity:
\begin{equation}
    \frac{\theta(u;\tau)}{\eta(\tau)^2}
    =u\,\exp\left(
        -2\sum_{k\geq 1}\frac{u^{2k}}{(2k)!}\,E_{2k}(\tau)
    \right)
\end{equation}
and the definitions of the Pontryagin classes in terms of the eigenvalues  $\pm 2\pi i\lambda_i$, $i=1,\ldots,\dim\mathcal{X}/2$ (when $\dim\mathcal{X}$ is odd, one of the eigenvalues is zero and the elliptic genus vanishes identically) of the curvature 2-form $\mathcal{R}$:
\begin{eqnarray}
        p_1 & = & \sum_{i}\lambda_i^2, \\
        p_2 & = & \sum_{i<j}\lambda_i^2\lambda_j^2, \\
        p_3 & = & \sum_{i<j<k}\lambda_i^2\lambda_j^2\lambda_k^2, \\
        \ldots 
\end{eqnarray}
We then have
\begin{multline}
	\int_\mathcal{X} \sqrt{\det\frac{\mathcal{R}}{\theta(\mathcal{R};\tau)}} =
	\frac{1}{\eta(\tau)^{\mathrm{dim}\mathcal{X}}} \int_\mathcal{X}
	\exp\left\{
	2\sum_{i}\sum_{k\geq 1}\frac{\lambda_i^{2k}}{(2k)!}\,E_{2k}(\tau)
	\right\}=\\
	\frac{1}{\eta(\tau)^{\mathrm{dim}\mathcal{X}}} \int_\mathcal{X}
	\exp\left\{p_1\,E_2(\tau)+\frac{p_1^2-2p_2}{12}\,E_4(\tau)+\frac{p_1^3-3p_1p_2+3p_3}{360}E_6(\tau)+\ldots
	\right\}
	=\\
	\frac{1}{\eta(\tau)^{\mathrm{dim}\mathcal{X}}} \int_\mathcal{X}
		\left\{
	1+E_2(\tau)p_1+\left(\frac{E_2(\tau)^2}{2}p_1^2+\frac{E_4(\tau)}{12}(p_1^2-2p_2)\right)+\right.\\
	\left.
	+\left(
	    \frac{E_2(\tau)^3}{6}\,p_1^3
	    +\frac{E_2(\tau)E_4(\tau)}{14}\,(p_1^3+2p_1p_2)+
	    \frac{E_6(\tau)}{360}\,(p_1^3-3p_1p_2+4p_3)+\ldots
	\right)
	\right\}
	\label{EG-deg-6}
\end{multline}
When $p_1=0$ the dependence on $E_2(\tau)$ drops out because it always comes multiplied with $p_1$, as it is clear from the second line of (\ref{EG-deg-6}).

\paragraph{Exercise \ref{ex:NS-flux}.} See Section 4.2 of \cite{Dabholkar:2020fde}.

\bibliographystyle{JHEP}
\bibliography{avatars}

\end{document}